\documentclass[preprint2,10pt]{aastex}
\usepackage{amsmath}

\def\teff{T$_{\rm eff}\,$}
\def\gsurf{g$_{\rm surf}\,$}
\def\tfall{$\tau_{\rm fall}\,$}
\def\vfall{$\rm v_{\rm fall}\,$}

\def\tcond{$\tau_{\rm cond}\,$}
\def\tcond{$\tau_{\rm cond}\,$}
\def\tconv{$\tau_{\rm conv}\,$}
\def\tcoag{$\tau_{\rm coag}\,$}

\def\tnuc{$\tau_{\rm nuc}\,$}
\def\smax{$\mathcal{S}_{\rm max}\,$}

\def\Dwa{$\,$\uppercase\expandafter{\romannumeral5}$\,$}

\def\sles{\lower2pt\hbox{$\buildrel {\scriptstyle <}
   \over {\scriptstyle\sim}$}}

\def\sgreat{\lower2pt\hbox{$\buildrel {\scriptstyle >}
   \over {\scriptstyle\sim}$}}
\def\sharpnull#1{}

\begin{document}

\slugcomment{\bf}
\slugcomment{Submitted to Ap.J.}

\title{Modeling the Formation of Clouds in Brown Dwarf Atmospheres}
\shorttitle{Brown Dwarf Clouds}

\author{Curtis S.\ Cooper\altaffilmark{1}, David Sudarsky\altaffilmark{2}, John
A. Milsom\altaffilmark{3}, Jonathan I. Lunine\altaffilmark{1}, Adam
Burrows\altaffilmark{2}}

\altaffiltext{1}{Department of Planetary Sciences and Lunar and Planetary
Laboratory, The University of Arizona, Tucson, AZ 85721;
                 curtis@lpl.arizona.edu, jlunine@lpl.arizona.edu}
\altaffiltext{2}{Department of Astronomy and Steward Observatory,
                 The University of Arizona, Tucson, AZ
		 85721; burrows@jupiter.as.arizona.edu, sudarsky@as.arizona.edu}
\altaffiltext{3}{Department of Physics, The University of Arizona, Tucson,
AZ 85721;  milsom@physics.arizona.edu}

\begin{abstract}
Because the opacity of clouds in substellar mass object (SMO) atmospheres
depends on the composition and distribution of particle sizes within the cloud,
a credible cloud model is essential for accurately modeling SMO spectra and
colors.  We present a one--dimensional model of cloud particle formation and
subsequent growth based on a consideration of basic cloud microphysics.  We
apply this microphysical cloud model to a set of synthetic brown dwarf
atmospheres spanning a broad range of surface gravities and effective
temperatures (\gsurf = $1.78\times 10^3$ -- $3\times 10^5$ cm s$^{-2}$ and
\teff = 600 -- 1600 K) to obtain plausible particle sizes for several
abundant species (Fe, Mg$_2$SiO$_4$, and Ca$_2$Al$_2$SiO$_7$).  At the base of the 
clouds, where the particles are largest, the particle sizes thus computed range from 
$\sim$$5\,\micron$ to over $300\,\micron$
in radius over the full range of atmospheric conditions considered.  We show
that average particle sizes decrease significantly with increasing brown dwarf
surface gravity.  We also find that brown dwarfs with higher effective
temperatures have characteristically larger cloud particles than those with
lower effective temperatures.  We therefore conclude that it is unrealistic when
modeling SMO spectra to apply a single particle size distribution to the entire
class of objects.
\label{abstract}
\end{abstract}


\section{Introduction}
\label{Introduction}

Substellar mass objects (SMOs) are fundamentally more complex than Sun-like stars because
of the formation of molecules in their cool outer layers.  These molecular
species cause the spectra to deviate strongly from blackbody values in a
multitude of spectral bands \citep[and references therein]{Leggett:1999,Kirkpatrick:1999,Burrows:1997}. 
Because of the complex chemistry occurring in their atmospheres, therefore, a 
comprehensive theory of SMOs requires detailed knowledge of the opacities of 
all the abundant molecular species.  Obtaining complete opacity data has been a 
major challenge for the field of SMO spectral synthesis \citep{Burrows:2001}.

The chemistry of brown dwarf and giant planet atmospheres is further complicated
by the condensation of gaseous molecules into liquid or solid cloud particles,
a process which occurs naturally at low temperatures.  Because clouds
significantly affect spectral features, a satisfactory theory for the structure
of substellar atmospheres must address cloud particle formation and subsequent
growth. Detailed knowledge of the distribution of particle sizes near the
photosphere is a basic requirement for properly modeling the optical effects of
clouds \citep{Lunine:1989,Ackerman:2001}.

Clouds influence brown dwarf and giant planet spectra in several important ways.
First, cloud formation can deplete the atmosphere of refractory elements that
become sequestered in condensed form and then rain out from the upper
atmosphere.  This effect is hypothesized to be important for the interpretation
of the L to T dwarf spectral transition \citep{Burrows:1999,Burrows:2001}.
Second, clouds near the photosphere will have the general
effect of smoothing out prominent spectral features \citep{Jones:1997}.
Third, the optical albedo of irradiated objects will be increased
substantially by the presence of clouds \citep{Sudarsky:2000}.  Fourth,
the presence of an optically thick cloud layer causes a back-warming effect 
that results in heating of the atmosphere. 

Clouds may be manifest in brown dwarf spectra in a more subtle way.
\citet{Bailer-Jones:2001} reported photometric I-band variability of up to
$7\%$ in L dwarf spectra.  The effect is a possible signature of variations due
to the patchiness of clouds.  Unfortunately, cloud patchiness in SMOs is not
well understood because even though much is known about the dynamic meteorology of the Earth,
these gaseous bodies are dynamically very different.  For example, general circulation models of the
Earth, although they are of great utility for studying Mars, completely fail to
reproduce the observed dynamics in Jupiter's atmosphere.
A detailed, three--dimensional cloud model paralleling the state of the art in Earth
climate models would require detailed meteorological measurements for
proper parameter calibration that are simply not available for any other planet.

The present cloud model at its core employs the basic microphysical timescale arguments
of \citet{Rossow:1978} to determine the most probable particle size in a cloud at
each vertical pressure level.  The model is one--dimensional only.  We make no
attempt to treat the intricacies of cloud patchiness or the effects of winds and horizontal advection.  
The present cloud model is not intended to calculate the detailed meteorology of
planetary atmospheres but to offer a simple prescription for estimating the opacity of clouds
in SMO atmospheres.  A detailed but computationally cumbersome SMO climate model, if it could be
developed, would be difficult to incorporate into spectral synthesis models.  The philosophy
behind a simple, one--dimensional approach to cloud modeling is to develop a
prescription based on well--established physical principles that can be used to
guide and inform spectral synthesis calculations.  The present model 
aims to characterize SMO cloud particle sizes and densities in a global and time averaged
sense.  We therefore assume the particle number density of the cloud to be uniform across the spherical surface 
of an object at a given pressure and temperature, and that there has been sufficient time in these systems to establish 
chemical equilibrium throughout the atmosphere.  

Our timescale arguments compute an average effectiveness for each of the various competing
physical processes.  Therefore, the output particle sizes clearly depend directly 
on our assumptions of the values of the four unknown parameters in the microphysical timescales
(see Section \ref{subsection:Cloud_Code}).  The calculation represents a 
first--order estimate of SMO cloud particle sizes, and we therefore present our results for SMO 
particle sizes with the realization that the procedure employed could potentially overestimate or 
underestimate the true particle sizes by some unknown factor, which we expect will be of order unity.  
Nevertheless, since we have been consistent in the assumed values of the unknown physical parameters 
throughout the calculation, we expect that the errors in the computed particle sizes will be uniform 
among the SMOs studied.  We are therefore confident, despite the inaccuracies of the approach, that the sizes computed
are correct to within an order of magnitude and that the \em trends \em 
we demonstrate, in which typical cloud particle sizes vary systematically with brown dwarf effective temperature and 
gravity, do represent physically meaningful results.

In this paper, we reiterate the general conclusion arrived at in \citet{Lunine:1989} 
and \citet{Ackerman:2001}:~it is not satisfactory
when modeling brown dwarf spectra to assume \em a priori \em a single particle size
distribution because the sizes of cloud particles vary strongly with effective
temperature, surface gravity, and height in the atmosphere.  We extend the previous efforts to incorporate
clouds into spectral models by calculating particle sizes based on a
one-dimensional model of cloud particle growth for several abundant species
over a range of realistic atmospheric conditions.  We compute particle radii
spanning a broad range from about $5\,\micron$ to over $300\,\micron$.
Therefore, in the context of this discussion, we hereafter refer to particles
less than $10\,\micron$ in radius as small; we refer to particles in the
range from $10-100\,\micron$ in radius as medium-sized; and we consider large
particles to be those having radii greater than $100\,\micron$.  

In Section \ref{section:Cloud_Model}, we present an improved model of cloud formation and droplet
growth to determine the composition, abundance, and distribution of cloud
particles in brown dwarf atmospheres.  In Section \ref{section:Model_Results}, we obtain 
modal cloud particle sizes for several representative cloud--forming species of high
abundance for atmospheric models spanning a broad range of effective
temperatures and surface gravities.  In Section \ref{section:Cloud_Model_Comparisons}, 
we compare our particle sizes with those of two other recent papers that have also addressed clouds 
\citep{Ackerman:2001,Helling:2001}. In Section
\ref{section:Discussion}, we apply our cloud model to brown dwarf spectra 
by computing the opacity of clouds resulting from the wavelength--dependent 
effects of Mie scattering and absorption of radiation. 
We also demonstrate the importance of clouds as an opacity source by
comparing the spectrum of a cloudy brown dwarf atmosphere with that of a
cloud--free atmosphere.

\section{The Cloud Model}
\label{section:Cloud_Model}

\subsection{Condensation Level}
\label{subsection:Condensation_Level}

The present cloud model follows much of the formalism of \citet{Lewis:1969}, 
\citet{Rossow:1978}, \citet{Stevenson:1988}, and \citet{Lunine:1989}.  To aid
the reader, we list in Table \ref{table:Model_Symbols} all the symbols used in our 
cloud model as described in this section of the text. 

\begin{deluxetable}{ll}
    \tablewidth{0pt}
    \tablecolumns{2}
    \tablecaption{Cloud Model Variables and Parameters \label{table:Model_Symbols}} 
    \tablehead{
    \colhead{Symbol}&\colhead{Description}\\}
    \startdata
$\rm g_{surf},\, T_{eff}$ & SMO surface gravity and effective temperature.	\\
$\rm P_{cond},\,P_{sat}$ & Partial pressure of the condensing vapor and its saturation vapor pressure.			\\
$\rm \mathcal{S}_{max},\,S$ & Assumed maximum supersaturation and the saturation ratio.	\\
$\rm \tau_{cond},\,\tau_{nuc}$ & Timescales for heterogeneous and homogeneous nucleation.			\\
$\rm \tau_{coal},\,\tau_{coag}$ & Timescales for coalescence and coagulation.	\\
$\rm \tau_{fall},\,\tau_{conv}$ & Timescales for gravitational fallout and convective upwelling.		\\
$\rm \tau_{rad}$    &	Timescale for particles to cool by radiation.		\\
T, P, $\rm \rho,\, \eta$ & Temperature, pressure, mass density, and dynamic viscosity of the atmosphere.		\\
$\rm R, k_b, N_A$ & Universal Gas Constant, Boltzmann's Constant, and Avogadro's Number.	\\
$\rm \mu, \,\mu_p$ & Mean molecular weight of atmosphere and condensate molecular weight.				\\
Kn, Re  &	Knudsen and Reynolds numbers.					\\
$\rm \lambda,\,c_{sound}$	&  Mean free path of atmospheric gas particles and atmospheric sound speed.		\\ 
v   &   Relative velocity between particle and fluid in Re expression.		\\
$\rm v_{fall}(r),\,g$	&  Terminal velocity of particles of radius r in local acceleration of gravity, g.		\\
$\rm v_{conv}$	& Upward velocity of convecting gas parcels.			\\
H   &	Atmospheric scale height; i.e., pressure e-folding distance.		\\
$\rm \epsilon_{surf}$, L	&   Surface tension of condensed molecules and the latent heat of vaporization.	\\
$\rm r_c$, Z   &	Critical radius of particles and the Zeldovich factor of classical nucleation theory.	\\
N	&   Number density of cloud particles.					\\
$\rm \epsilon_{coal},\,\epsilon_{coag}$	& Coalescence and coagulation efficiencies.				\\
$\rm q_c$   &	Moles of condensate per mole of atmosphere.			\\
$\rm q_v$   &	Moles of condensable vapor per mole of atmosphere.		\\
$\rm q_t$   &	Total moles of condensable material per mole of atmosphere.	\\
$\rm q_{below}$	&  Mole fraction of condensable vapor below the cloud base.	\\  
$\rm P_{c,1}$	&  Condensation curves as shown in Figure \ref{figure:Condensation_Level}.	\\
$\rm n(r),\, r_0$  &	Particle size distribution about modal particle size, $\rm r_0$.	\\

    \enddata
\end{deluxetable}

As in \citet{Lewis:1969}, we compare the partial pressure of a given condensable species to
its saturation vapor pressure.  Only at levels where the partial pressure
exceeds the saturation vapor pressure can condensation begin.  However, the
required level of supersaturation for efficient condensation depends on the
physics of the nucleation process itself.  We discuss the two relevant
nucleation processes---homogeneous and heterogeneous nucleation---in more detail
in Section \ref{subsection:Microphysical_Timescales}.  In a clean atmosphere, 
in which the availability of condensation nuclei is low (the case of homogeneous nucleation), 
condensation will not begin until the vapor becomes highly supersaturated.  Thus, the
partial pressure will greatly exceed---often by a factor of two or more---the
saturation vapor pressure \citep{Rogers:1989}.  This situation is hypothesized to
occur in methane clouds in Titan's atmosphere \citep{Tokano:2001}.
The condensation level appears where
\begin{equation} \label{equation:Condensation_Level}
\textrm {P}_{\rm cond} >
\textrm {P}_{\rm sat} \, (1+\mathcal{S}_{\rm max}).
\end{equation}
In Equation \ref{equation:Condensation_Level}, $\textrm {P}_{\rm cond}$ is the partial pressure 
of the condensing vapor, and $\textrm {P}_{\rm sat}$ is its saturation vapor pressure.  
The \smax parameter is the assumed maximum supersaturation.  For example, 
if $\rm \mathcal{S}_{max} = 0.01$, then the partial pressure 
of the condensing vapor must be $\rm 1\%$ larger than the saturation vapor pressure before
condensation can begin.  To aid future discussions, we also define the \em saturation ratio \em as $\rm S =
P_{cond}/P_{sat}$.

In addition to the original \citet{Lewis:1969} cloud model, we also treat the case
of heterogeneous (or chemical) condensation, in which chemical constituents
different in composition from the cloud react chemically to produce the
condensate.  For example, we allow Mg, Si, and O to produce forsterite,
Mg$_2$SiO$_4$.  In such cases of heterogeneous condensation, the concept of a
saturation vapor pressure is replaced by the heterogeneous condensation
pressure.  We apply the chemical equilibrium model of \citet{Burrows:1999}
to a solar--composition gas \citep{Anders:1989} to determine which chemical 
species are thermodynamically favored. The calculation determines the equilibrium compositions
by minimizing the Gibbs free energy of the system.

The maximum supersaturation, $\rm \mathcal{S}_{max}$ 
in Equation \ref{equation:Condensation_Level}, appears in the calculation as an essentially free 
parameter, though we have some guidance as to its likely value from measurements of the 
supersaturation of water clouds on Earth.  We assume in these calculations that the abundance of condensation
nuclei is sufficient for heterogeneous nucleation to begin at only $1\%$ maximum
supersaturation, though we will explore the effect of this parameter on the
computed particle sizes subsequently (see Section \ref{subsection:Varying_Free_Parameters}).  We 
have not attempted herein to calculate the precise composition or abundance of
these condensation nuclei.  But even for iron, which is the most refractory of the homogeneously
condensing species, we argue based on chemical equilibrium considerations  
that there exist abundant refractory condensates that can potentially serve as nucleation sites for
heterogeneous nucleation, including corundum ($\rm Al_2O_3$), hibonite ($\rm CaAl_{12}O_{19}$), the calcium
titanates (e.g., $\rm CaTiO_3$ and $\rm Ca_4Ti_3O_{10}$), 
grossite ($\rm CaAl_4O_7$) and other calcium-aluminum oxides, and spinel ($\rm MgAl_2O_4$).  

As in \citet{Rossow:1978}, we ignore variations in $\rm \mathcal{S}_{max}$ within the cloud.  Although in reality
the supersaturation should decrease with altitude, we ignore this effect in
the present calculations.  Furthermore, it is possible to have a small amount
of condensate present between $\rm S = 1$ and $\rm S = 1 + \mathcal{S}_{max}$,
even though condensate cannot possibly form there (since the initial condensation of vapor into
embryonic particles in general requires at least a modest level of supersaturation).  However,
particles can settle into the region between $\rm S = 1$ and $\rm S = 1 + \mathcal{S}_{max}$
once they have formed and yet still be thermodynamically stable.  Since it depends on a
balance between the processes of gravitational settling and convective upwelling, the amount of
cloud material present in the barely supersaturated region is difficult to calculate with
certainty.  The clouds we treat in this article have a very low $\rm \mathcal{S}_{max}$, 
and therefore the errors introduced by neglecting these effects are small.  In a cloud having very
few condensation nuclei, which as we have stated would require a large $\rm \mathcal{S}_{max}$,
considerations such as the variation of the supersaturation with altitude and
the exact location of the cloud base become more crucial for accurate cloud
modeling.  The present cloud model is not configured to account for these
effects realistically, though they are relatively unimportant for the low supersaturations
generally assumed throughout this paper.

\begin{figure}
\includegraphics[height=3.7in, width=2.8in, angle=-90]{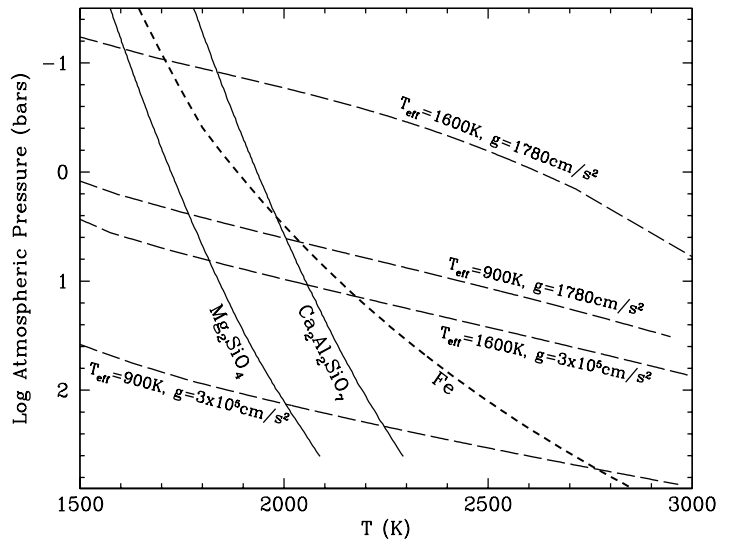}
\caption{
Illustrates the locations of the cloud bases for four different atmospheric
models (long--dashed curves) and three different chemical species (solid and
short--dash curves). The atmospheric models are at two different surface
gravities and two different effective temperatures.  The short--dashed curve
defines the atmospheric pressures and temperatures for which the vapor pressure
of iron equals its saturation vapor pressure.  This curve is obtained using the
iron saturation vapor pressure expression in \citet{Barshay:1976}.  The saturation
vapor pressure curve has been adjusted using the iron vapor mixing ratio (by
number) below the cloud (parameter $\rm q_{below}$ in 
Table \ref{table:Condensate_Properties}).  The two solid lines are the curves which
illustrate when condensation of forsterite ($\textrm{Mg}_2\textrm{SiO}_4$) and
gehlenite ($\textrm{Ca}_2\textrm{Al}_2\textrm{SiO}_7$) become
thermodynamically favored. Thus, they represent the pressure dependent condensation temperatures
of those species under the assumption that chemical equilibrium holds.  The
intersection of a given atmospheric model and the two condensation curves
represents the pressures and temperatures at the bases of the chemically
condensing clouds in our cloud model. However, for iron, the cloud base will be
at lower temperature and pressure than the intersection point if the iron vapor is significantly
supersaturated (see Equation \ref{equation:Condensation_Level}).  
}
\label{figure:Condensation_Level}
\end{figure}

Figure \ref{figure:Condensation_Level} shows the condensation curve for two condensates, forsterite
(Mg$_2$SiO$_4$) and gehlenite (Ca$_2$Al$_2$SiO$_7$), as well as the ``effective
condensation curve'' of iron derived from the iron saturation
vapor pressure curve, and several model brown dwarf atmospheres.
As the caption to Figure \ref{figure:Condensation_Level} explains, the curve labeled ``Fe'' is not
the true saturation vapor pressure curve of iron.  It is an ``effective condensation
curve'' derived by taking the saturation vapor pressure of iron and dividing by
the number mixing ratio of iron vapor just below the cloud base (parameter $\rm q_{below}$ 
of Table \ref{table:Condensate_Properties}), which is the maximum
allowed number mixing ratio of gaseous iron in a solar composition mixture when hydrogen is all in
molecular form ($\rm H_2$).  We discuss this further in Section \ref{subsection:Cloud_Code}.

The intersection points between the four long--dashed curves of brown dwarf
temperature--pressure profiles and the condensation curves of forsterite and
gehlenite represent levels in the atmosphere at which the relevant species may
appear via one or more chemical reactions.

The more familiar case of homogeneous condensation, or direct condensation
from a vapor of the same composition, does occur for water and iron.  Thus, for
the condensation of iron and water, we follow \citet{Lewis:1969} and employ the
well--known saturation vapor pressure relations of iron and water \citep{Barshay:1976,Lunine:1989}.
In Figure \ref{figure:Condensation_Level}, 
the iron curve is thus shown as a thick, short--dashed line as a reminder that the iron cloud condenses
directly from iron vapor.  As we discuss in more detail in Section
\ref{subsection:Coupling_Clouds_With_SMO_Atmosphere_Models}, in which we
explore the coupling between clouds and the radiative transfer problem, the atmospheres
used in this stage are computed on the basis of an independent, dust--free
stellar atmosphere code.

\subsection{Microphysical Timescales}
\label{subsection:Microphysical_Timescales}

Although the chemistry applying to brown dwarf atmospheres is radically
different from Earth's atmospheric chemistry, we nevertheless treat the
various competing effects in the cloud formation process as microphysical
timescales \citep{Rossow:1978}.  In our model, five microphysical processes compete 
with gravitational fallout to increase the modal particle size:~1) convective
uplifting of condensable vapor, 2) heterogeneous nucleation, 3) homogeneous
nucleation, 4) coagulation, and 5) coalescence.  Each process is characterized
by a single-valued timescale which expresses its relative importance.

The expressions for four of the microphysical timescales are computed in
\citet{Rossow:1978}: $\tau_{\rm cond}$ (heterogeneous condensation), $\tau_{\rm coal}$
(coalescence), $\tau_{\rm coag}$ (coagulation), and $\tau_{\rm fall}$ (gravitational fallout).  
These expressions depend on the values of both
the Knudsen and Reynolds numbers, which characterize how particles of different sizes interact
physically with the surrounding fluid and with each other.  
It is therefore necessary to calculate these values during the particle 
growth phase of the calculation (see Section \ref{subsection:Cloud_Code}).  

We also require an equation of state relating the gas density 
to its temperature and pressure.  In this model, we employ the ideal gas equation of state 
to relate the gas mass density $\rm \rho$ to the atmospheric temperature and pressure (T, P): 
$\rm \rho = \mu P/RT$, where R is the universal gas constant, $\rm 8.314\times
10^7\,ergs\,mol^{-1}\,K^{-1}$.
The equation of state depends on the mean molecular weight of the atmospheric
gas mixture.  For these atmospheres, we use $\rm \mu = 2.36\;g\,mol^{-1}$, the mean molecular weight of a solar
composition gas in which hydrogen is present in the molecular state, 
$\rm H_2$.  We have used the chemical equilibrium model of \citet{Burrows:1999} to verify 
that---as for Jupiter---molecular hydrogen is the dominant phase of hydrogen in the upper atmospheres of SMOs.  
The ideal gas equation of state reproduces to within 1\% the value of $\rm \rho$ obtained 
using the more sophisticated thermodynamic equation of state described in \citet{Burrows:1989}.

The dimensionless Knudsen number differentiates between the two
regimes---classical and gas kinetic---of physical interaction between cloud 
particles and the surrounding medium:  
\begin{equation}
\rm Kn = \lambda/r, 
\label{equation:Knudsen_Number}
\end{equation}
where $\rm r$ is the cloud particle's radius and $\lambda$ is the mean free path of 
atmospheric gas particles. The transition between the two regimes occurs where
the Knudsen number is near unity (i.e., where $\rm \lambda \sim r$).  The high 
Knudsen number regime, in which the mean free path of gas molecules is much larger than the size of a 
typical cloud particle, will normally not occur in SMO clouds.  However, there are 
two situations in which large Knudsen numbers could arise: 1) when particles are smaller than
$\sim$$1\,\micron$ in radius, and 2) high up in the atmosphere of a planet or brown dwarf at pressures
much less a bar.  We include the Knudsen number in the cloud model to account for these
possibilities.
  
The dimensionless Reynolds number characterizes the behavior of a spherical particle moving
through a fluid.  The equation for the Reynolds number is
\begin{equation}
\rm Re = \frac{2\rho vr}{\eta},
\label{equation:Reynolds_Number}
\end{equation}
where $\rm \rho$ is the mass density of gas in the atmosphere, $\rm r$ is the radius of 
the particle, v is the velocity of the particle relative to the fluid, and 
$\eta$ is the dynamic viscosity of the fluid.  The flow will be turbulent 
for $\rm Re\gg10$ but laminar for low Reynolds numbers \citep{Landau:1959}.  
For the purposes of these calculations, we assume laminar flow unless 
$\rm Re > 70$ (see Section \ref{subsection:Cloud_Code}).  
This choice for the laminar to turbulent cutoff is convenient 
because above $\rm Re = 70$, the drag coefficient describing the variation in terminal 
velocity of a falling particle becomes roughly a constant at $\rm C_D\approx 0.2$
\citep{Rossow:1978}. 

To calculate the Reynold's number and the timescale for gravitational fallout, 
we introduce an approximate value for $\eta$, the dynamic viscosity.  
The viscosity of a hydrogen and helium gas mixture at low densities is roughly 
constant over a broad range of temperatures.  Because the viscosity increases
with the square root of temperature at low pressures, it varies by less than a factor 
of two over the full range of SMO effective temperatures discussed herein
(see Section \ref{subsection:Cloud_Code}).  For simplicity, our model therefore assumes a 
constant value for the dynamic viscosity of $\rm \eta=2\cdot10^{-4}\,poise$, 
which is consistent with the measured values of $\eta$ for hydrogen and helium 
at low densities and $\rm T\sim 1000\,K$ \citep[CRC Handbook]{CRC:1991}.

The \tfall timescale represents gravitational fallout of particles, which
is the fundamental process limiting particle growth.  As the particles become
large, they fall out.  We take the gravitational fallout timescale to be the
time required for particles to fall through a pressure scale height of the
atmosphere at terminal velocity.  

The terminal velocity of falling particles depends on their size, shape,
density, rigidity, and the nature of the fluid flow around them
(e.g., Stokes flow, free molecular flow, or turbulent flow).  We assume rigid spherical particles 
throughout this paper.  While it is true that liquid particles of a given size 
will fall more slowly through the atmosphere than solid particles because 
of the disruption of their shape, the effect will not 
significantly alter the fall speeds for particles smaller than about a 
millimeter \citep{Rogers:1989}.  The particles we obtain using this model---see Section 
\ref{section:Model_Results}---never grow large enough 
for the difference in terminal velocity between solid and liquid particles 
to be noticeable.  Thus, taking the particles to be 
rigid spheres is in this case a reasonable simplification.

For the common case of low Knudsen and low Reynolds number flow, 
the terminal velocity will be given by the Stokes solution for a falling rigid 
sphere \citep{Rogers:1989}.  The Cunningham factor, $\rm 1 + \case{4}{3}$Kn, 
accounts for the variation in the terminal velocity with Knudsen number.  
There is also a variation in terminal velocity with Reynold's number.  
These combined effects yield three different forms for the terminal velocity 
\vfall as a function of the radius r of the particle:
\begin{equation}
\rm v_{fall}(r) = r^2\cdot\left( \frac{2\rho_{cond}g}{9\eta} \right) \;\;\;\;\;Re < 70,\;Kn < 1,
\label{equation:Stokes_Terminal_Velocity}
\end{equation}
\begin{equation}
\rm v_{fall}(r) = r^{1/2}\cdot\sqrt{\frac{40\rho_{cond} g}{3\rho} }
		    \;\;\;\;\;Re > 70,\;Kn < 1,\;and
\label{equation:Turbulent_Terminal_Velocity}
\end{equation}
\begin{equation}
\rm v_{fall}(r) = r\cdot\sqrt{\frac{\pi \mu}{2N_A k_b T}}\,
		\left(\frac{8 \rho_{cond} g}{27 \rho}\right) \;\;\;\;\;Kn > 1.
\label{equation:Gas_Kinetic_Terminal_Velocity}
\end{equation}

In Equations \ref{equation:Stokes_Terminal_Velocity}, 
\ref{equation:Turbulent_Terminal_Velocity}, and \ref{equation:Gas_Kinetic_Terminal_Velocity}, 
$\rm N_A$ is Avogadro's number ($\rm 6.022\times 10^{23}\,mol^{-1}$), $\rm k_b$ is Boltzmann's constant 
($\rm 1.38\times 10^{-16}\,erg\,K^{-1}$), and g is 
the local gravitational acceleration.  The quantity $\rm \rho_{cond}$ represents 
the mass density of the particle itself, for which we use the bulk density of a solid or 
liquid phase having the composition of the condensate.  The parameters used for each 
condensate we have treated---iron, forsterite, gehlenite, and water---are listed in Table
\ref{table:Condensate_Properties}.  The mass density of the gas 
$\rm \rho$ depends on temperature and pressure through the 
equation of state.  The terminal velocity of particles depends also on the local 
acceleration of gravity, $\rm g$, which varies considerably among the SMO population (see Section
\ref{section:Model_Results}).  In the Stokes solution, 
Equation \ref{equation:Stokes_Terminal_Velocity},    
the terminal velocity is proportional to the square of
the radius of the falling sphere.  In the turbulent regime,
Equation \ref{equation:Turbulent_Terminal_Velocity}, it 
is proportional to the square root of the radius.  In the gas
kinetic (or large Knudsen number) regime,
Equation \ref{equation:Gas_Kinetic_Terminal_Velocity}, the terminal velocity 
increases linearly with particle radius.  The gravitational fallout timescale
is inversely proportional to the terminal velocity: $\rm \tau_{fall} =
H/v_{fall}$, where $\rm H = \case{\rm RT}{\rm \mu g}$ is the atmospheric scale height.  
We describe in Section \ref{subsection:Cloud_Code} our procedure for
differentiating between the Reynolds number regimes, which is necessary to calculate
\tfall and the other microphysical timescales.

The \tconv timescale represents the convective upwelling of particles.
Although convective upwelling is not itself a particle growth process,
convective updrafts hold small cloud particles aloft in the atmosphere so that
they accumulate more material via coagulation, coalescence, etc.  We define
the timescale characterizing convective updrafts as \tconv = $\rm H/\rm v_{\rm conv}$.  
The convective timescale is thus the time required for gas parcels 
to rise through a pressure scale height. The upward convective velocity, $\rm v_{conv}$, 
is estimated via a mixing length prescription in the atmosphere code.

The nucleation timescale, $\rm \tau_{nuc}$, characterizes particle growth via homogeneous nucleation, in
which highly supersaturated vapor condenses spontaneously due to rapid
collisions between the vapor molecules.  For the nucleation timescale, we
employ the classical homogeneous nucleation theory, in which droplets form in free space 
as a result of chance collisions between the molecules of the supersaturated vapor.  
This process is rarely observed to occur in nature because its timescale will be 
longer in almost all cases than \tcond, which characterizes the condensation growth of a population
of particles via heterogeneous nucleation.  Condensation
growth means the conversion of vapor to particles, whether they be in the liquid or solid phase.  
For example, the phase diagram of iron indicates that iron will condense into liquid particles 
above 1800 K.  However, we ignore the distinction between liquid and solid particles because the
particles are small enough that the flattening effect during free--fall is
negligible.  In heterogeneous nucleation, particles nucleate onto seed particles, 
called condensation nuclei, which are presumed to populate the region of condensation. 
As discussed, when heterogeneous nucleation is efficient, condensation may begin even when the
level of supersaturation is quite low ($<1\%$).  Neither nucleation process
represented by \tnuc and \tcond apply to the species that appear due to chemical reactions (e.g., forsterite and
gehlenite) rather than from a supersaturated vapor. The nucleation timescale expressions 
apply only to the species (e.g., iron and water) for which the condensate forms
directly from a vapor of the same composition.

For the condensation timescale, \tcond, we consider nucleation onto seed aerosols, 
also known as condensation nuclei.  The assumption of a low maximum supersaturation of 
the condensable vapor---$\rm \mathcal{S}_{max} = 1\%$ throughout (see Section
\ref{subsection:Condensation_Level})---is equivalent to the assumption that
condensation nuclei are abundant in the upper atmosphere; i.e., these are 
``dirty'' atmospheres.  We have not attempted in the present model to calculate the
compositions or exact abundances of these condensation nuclei.  Our assumed value 
for \smax is based rather on the recognition that the complex chemistry 
occurring in SMO atmospheres will likely generate a plethora of molecules 
suitable as seeds for the onset of heterogeneous nucleation.  The expressions for the
condensation timescale, which depend on the Knudsen number regime, 
are adapted from \citet{Rossow:1978}:
\begin{equation}
\rm \tau_{cond} = \rho r^2\left(\frac{\rho_{cond}RT}{4\eta
\mu P_{sat}\mathcal{S}_{max}}\right),\;\;\;\;\; Kn < 1,
\label{equation:Condensation_Timescale_Classical_Gas}
\end{equation}
\begin{equation}
\rm \tau_{cond} = r\left(\frac{\rho_{cond}RT}{3P_{sat}\mathcal{S}_{max}}\right)
	    \left(\frac{\pi}{2\mu N_A k_b T}\right)^{1/2},\;\;\;\;\; Kn > 1.
\label{equation:Condensation_Timescale_Gas_Kinetic}
\end{equation}
In Equations \ref{equation:Condensation_Timescale_Classical_Gas} and 
\ref{equation:Condensation_Timescale_Gas_Kinetic}, $\rm P_{sat}$ is the 
saturation vapor pressure defined in Equation \ref{equation:Condensation_Level}, 
and $\rm \rho_{cond}$ is the mass density of the condensate.  
The other variables are the same as in Equations
\ref{equation:Stokes_Terminal_Velocity},
\ref{equation:Turbulent_Terminal_Velocity}, and
\ref{equation:Gas_Kinetic_Terminal_Velocity}.  
The mass densities of the four condensates treated in this paper are given in Table~1.

The condensation timescale given in Equations
\ref{equation:Condensation_Timescale_Classical_Gas} and
\ref{equation:Condensation_Timescale_Gas_Kinetic} assumes that 
heterogeneous nucleation dominates over homogeneous nucleation; i.e. $\rm \mathcal{S}_{max}\ll 1$.  
We relax this assumption by incorporating the process
of homogeneous nucleation explicitly ($\rm \tau_{nuc}$).  It should be noted,
however, that the inclusion of $\rm \tau_{nuc}$ will only be important if the
maximum supersaturation parameter is assumed to be large ($\rm \mathcal{S}_{max}\gg 1$).  We 
present homogeneous nucleation as a feature of the cloud model so that particle growth can be
treated in the extreme case of very high supersaturations.  As we have
indicated, however, this situation is not likely in brown dwarf atmospheres.
In most cases, homogeneous nucleation will be orders of magnitude slower than
condensation growth by heterogeneous nucleation (i.e., $\rm \tau^{-1}_{nuc} \ll 
\tau^{-1}_{cond}$).

For the timescale of homogeneous nucleation, we have adapted expressions from \citet{Stevenson:1988} 
for large Knudsen numbers:
\begin{equation}
\rm \tau_{\rm nuc}= \frac{\rho_{cond}rL^2\mu}{c_{sound}RTP\left
[exp\left (2\epsilon_{surf}\mu/RT\rho_{cond}r\right) - 1\right ]}
\;\; (\textrm {Kn} > 1).
\label{equation:tau_nuc1}
\end{equation}
Here, $\rm r$ is the particle radius, $\rm L$ is the latent heat of vaporization, 
$\rm \epsilon_{surf}$ is the surface tension of the condensed liquid molecules, 
$\rm c_{sound} = \sqrt{RT/\mu}$ is the sound speed in the atmosphere, and P and T are the 
ambient temperature and pressure.  We list the values of these parameters for
iron and water in Table \ref{table:Condensate_Properties}.  This expression is derived from Equations 3-5 of 
\citet{Stevenson:1988} by taking $\rm \tau_{nuc} = r\left[\frac{dr}{dt}\right]^{-1}$ and then substituting 
in from the ideal gas equation of state and the 
Clausius-Clapeyron equation to eliminate the particle number density, saturation vapor pressure, 
and the vapor pressure temperature gradient.  We take the efficiency of heat
exchange in Equation 3 of \citet{Stevenson:1988} to be $\rm 100\%$ and the average relative velocity of 
cloud particles colliding with the local hydrogen to be $\rm c_{sound}$.

The $\rm \tau_{nuc}$ expression for large Knudsen numbers represents the timescale
for the growth of an embryonic cloud particle that must release the latent heat of
condensation through collisions with the surrounding hydrogen
\citep{Stevenson:1988}.  It is possible, however, for particles in high-temperature environments to
grow by releasing the latent heat of condensation simply through radiatively cooling.  
We therefore include in the model an estimate of the
radiative cooling timescale, which is compared with Equation \ref{equation:tau_nuc1} 
in the calculation.  If $\rm \tau^{-1}_{rad} > \tau^{-1}_{nuc}$,
and the mean free path of particles is greater than the particle size, we use
$\rm \tau_{rad}$ in place of $\rm \tau_{nuc}$.  The equation we use to
approximate $\rm \tau_{rad}$ is adapted from \citet{Woitke:1999}:
\begin{eqnarray}
\rm \tau_{rad} \approx \left(2\times10^{-2}\;sec\right)\cdot\left(\frac{\rho_{cond}}{1\;g\,cm^{-3}}\right)\times \nonumber\\
\left(\frac{\mu_p}{1\;g\,mol^{-1}}\right)^{-1}\left(\frac{T}{1000\;K}\right)^{-4}.
\label{equation:tau_rad}
\end{eqnarray}
In Equation \ref{equation:tau_rad}, $\rm \mu_p$ is the molecular weight of the condensate (as distinct from 
$\rm \mu\approx 2.36\;g\,mol^{-1}$, the mean molecular weight of the atmospheric gas mixture).  
The values of $\rm \mu_p$ for the four condensates treated herein
are given in Table \ref{table:Condensate_Properties}.  

It should be noted that in a very high temperature, low density environment, in which
the release of latent heat is extremely fast, the limiting timescale for
homogeneous nucleation will be neither $\rm \tau_{nuc}$ nor $\rm \tau_{rad}$ but the timescale for the
diffusion of vapor molecules to the surface of the grain.  We do not calculate
this timescale in the present model.  For the purposes of
SMO atmospheres, however, this regime will never be realized in practice because the
condensation temperatures of even the most refractory compounds rarely exceed
2000 K.  Therefore, although the limit of $\rm \tau_{rad} \to 0$ in this formulation
appears to be potentially problematic, the temperature constraints 
ensure that $\rm \tau_{rad}$ cannot be arbitrarily small.  
Hence, for use in the calculation of SMO clouds, it is reasonble to 
assume that the shorter of the two timescales, 
$\rm \tau_{nuc}$ and $\rm \tau_{rad}$, determines the particle's homogeneous
nucleation rate.  

The constant preceding Equation \ref{equation:tau_rad} is
strictly correct only for highly opaque species having high extinction efficiencies.  
This is appropriate for the condensates we focus on here, although the timescale for radiative cooling 
will be longer by a factor of ten or a hundred than the timescale shown 
in Equation \ref{equation:tau_rad} if the particles are mostly transparent.  
In cases in which the $\rm \tau_{rad}$ timescale dominates the particle
sizes, a more accurate approximation than we have made in Equation
\ref{equation:tau_rad} should be calculated by integrating the extinction
efficiency over wavelength; e.g., see Equation 17 of \citet{Woitke:1999}.

The more common case of small Knudsen numbers, which corresponds to high
atmospheric gas density, is a commonly used meteorological expression; see
\citet{Rogers:1989}.  The low Knudsen number expression for \tnuc
is written in terms of the critical radius at which the equilibrium vapor
pressure over the surface of a particle equals the ambient vapor
pressure.  The equilibrium vapor pressure over the surface of a liquid
particle, because of its finite curvature, is in general higher than the
equilibrium vapor pressure over a flat surface of the liquid, which is the
saturation vapor pressure normally measured in the laboratory.  Therefore, the
critical radius of the liquid particle will only be attained when the saturation
ratio exceeds unity; i.e., when the actual ambient pressure exceeds the
saturation vapor pressure.  The critical radius therefore depends on the
supersaturation \smax \citep{Rogers:1989}:
\begin{equation} 
\rm r_c = \frac{2 \mu_p\epsilon_{surf}}{\rho_{cond} R T \,
\textrm{ln} (1 + \mathcal{S}_{\rm max})}.
\label{equation:r_crit}
\end{equation}

Unlike the gas kinetic regime, in which the removal of latent heat limits the rate at which particles
can grow by nucleation, in the classical gas regime, the nucleation rate of particles is
determined by the rate at which supercritical droplets are formed; i.e., droplets
larger than the critical radius above which particles grow spontaneously.  
The expression for $\rm \tau_{nuc}$ in terms of 
the critical droplet radius $\rm r_c$, adapted from \citet{Rogers:1989}, reads
\begin{equation}
\rm \tau_{\rm nuc} = \frac{1}{P}\,\sqrt{ \frac{\mu_pk_bT}{8\pi N_A r_c^4} }
{\textrm{Z}}^{-1}\;exp\left(\frac{4\pi
r_c^2\epsilon_{surf}}{3k_bT}\right)\;\;(\textrm {Kn} < 1).
\label{equation:tau_nuc2}
\end{equation}
In Equation \ref{equation:tau_nuc2}, 
Z is the dimensionless Zeldovich or non--equilibrium factor, which depends on
temperature and the physical properties of the condensate.  The equation for
Z, given by \citet{Jacobson:1999}, is 
\begin{equation}
\rm Z = \frac{\mu_p}{2\pi r_c^2 \rho_{cond} N_A} \, \sqrt{\frac{\epsilon_{surf}}{k_b T}}.
\label{equation:Zeldovich_Factor}
\end{equation}
The Zeldovich factor accounts for the differences between equilibrium and
non--equilibrium cluster concentrations in the classical homogeneous nucleation 
theory used to derive Equations \ref{equation:r_crit} and \ref{equation:tau_nuc2}.

Coagulation and coalescence represent particle growth due to collisions.
The coagulation timescale, $\rm \tau_{coag}$, refers to the formation of larger particles by the collision
of smaller particles.  It therefore depends primarily on the thermal temperature,
viscosity, and the number density of cloud particles. The coalescence
timescale, $\rm \tau_{coal}$, characterizes the growth of particles by coalescence, in which large particles
having high downward velocities in the fluid overtake and merge with small,
slowly falling particles. Thus, coalescence refers to the collisional process
caused by the different fall speeds of different sized particles, whereas
coagulation is collisional growth resulting from Brownian motion.  Both
processes proceed at a rate proportional to the particle number
density, N, since the total amount of material available for condensation is conserved.  
The number density of cloud particles is given in terms of the gas density as 
\begin{equation}
\rm N = q_c\left(\frac{\rho}{\mu}\right)\left(\frac{3\mu_p}{4\pi\rho_{cond}r^3}\right).
\label{equation:Particle_Number_Density}
\end{equation}
In Equation \ref{equation:Particle_Number_Density}, r is the cloud particle
radius, $\rm \rho$ is the mass density of the surrounding gas (as given by the
equation of state), $\rm \rho_{cond}$ is the mass density of the condensate, and $\rm q_c$ is the
condensate mixing ratio by number, which depends on the specific rainout
prescription applied.  The rainout scheme we employ to compute the variation
of $\rm q_c$ with height above the cloud base is explained in Section \ref{subsection:Cloud_Code}.

Given the particle number density, N, the timescale for coagulation in terms
of the radius of the condensed particle, $\rm r$, is given by
\begin{equation}
\rm \tau_{coag} \equiv \frac{N}{dN/dt}=
\frac{3\eta}{4k_bT\epsilon_{coag}}\,\frac{1}{N},\;\;\;\;\;Kn<1
\label{equation:Coagulation_Timescale_Classical_Gas}
\end{equation}
\begin{equation}
\rm \tau_{coag} = \frac{1}{4\epsilon_{coag}}\sqrt{\frac{\rho_{cond}}{3rk_bT}}\,\frac{1}{N},\;\;\;\;\;Kn>1.
\label{equation:Coagulation_Timescale_Gas_Kinetic}
\end{equation}
Here, $\rm \epsilon_{coag}$ is the coagulation efficiency, which is a free
parameter of the model.  We discuss this further in Section
\ref{subsection:Free_Parameters}.

Our final microphysical timescale, $\rm \tau_{coal}$, characterizes
the process of coalescence, in which larger particles overtake and coalesce
with smaller particles.  For coalescence, the Reynolds number again plays an
important role because---unlike coagulation---coalescence pertains specifically to the merging
of particles of greatly different radii whose interaction depends on the
nature of the surrounding flow fields.  The coalescence expressions are adapted from \citet{Rossow:1978},
although we have introduced an extra free parameter into the equations, 
the efficiency for coalescence, $\rm \epsilon_{coal}$:
\begin{equation}
\rm \tau_{coal} =
\frac{1}{r^4\epsilon_{coal}}\left(\frac{9\eta}{\pi\rho_{cond}g}\right)\,\frac{1}{N},\;\;\;\;\;Re<70, Kn<1
\label{equation:Coalescence_Timescale_Laminar}
\end{equation}
\begin{equation}
\rm \tau_{coal} = \frac{1}{\epsilon_{coal}}\sqrt{\frac{3\rho}{10\pi^2\rho_{cond} r^5 g}}\,\frac{1}{N},
\;\;\;\;\;Re>70, Kn<1
\label{equation:Coalescence_Timescale_Turbulent}
\end{equation}
\begin{equation}
\rm \tau_{coal} = 
\frac{1}{r^3\epsilon_{coal}}\left(\frac{27\rho}{4\pi\rho_{cond}g}\right)
\sqrt{\frac{2RT}{\pi \mu}}\,\frac{1}{N},\;\;\;\;\;Kn>1.
\label{equation:Coalescence_Timescale_Gas_Kinetic}
\end{equation}

As we discuss in Section \ref{subsection:Free_Parameters}, coalescence is 
important only for liquid particles.  This is the only case in our model in which
the distinction between liquid and solid particle formation is important.
Thus, for iron cloud formation above 1800 K, coalescence will potentially be important, and it 
can be important for water clouds, but we ignore it for the silicate clouds.

\subsection{Our Cloud Code}
\label{subsection:Cloud_Code}

The code proceeds in two stages: (1) determination of the cloud base
globally in the atmosphere, and then (2) particle growth at each atmospheric
temperature--pressure level.  The algorithm is run for each of the
atmospheric temperature--pressure profiles, which differ in effective temperature
and gravity, and for each condensable species separately.  In our model, we
ignore \em core--mantle \em grains composed of molecules of two or more of the major species.  
These types of condensates could potentially be important in regions in which
several major constituents in the equilibrium vapor mixture form at similar temperatures
and pressures.  In the present cloud model, we do not treat the growth of grains composed of 
multiple chemical phases.

We list in Table \ref{table:Condensate_Properties} the input parameters for the four condensable 
species treated herein. 

\begin{deluxetable}{lcccc}
 \tablewidth{0pt}
 \tablecolumns{5}
 \tablecaption{Condensate Physical Properties\label{table:Condensate_Properties}} 
 \tablehead{
 \colhead{Model Parameter}&\colhead{Iron}&
\colhead{Forsterite}&\colhead{Gehlenite}&\colhead{Water}\\}
 \startdata
Molecular weight, $\rm \mu_p$ [$\rm g\,mol^{-1}$] & 55.8 & 140.7 & 274.2 & 18.0 \\
Mass Density, $\rm \rho_{cond}$ [$\rm g\,cm^{-3}$] & 7.9\tablenotemark{a} & 3.2 & 3.0 & 1.0\tablenotemark{a} \\
Mole Fraction, $\rm q_{below}$\tablenotemark{b} & $5.4\times 10^{-5}$ & $3.2\times 10^{-5}$ & $1.8\times
10^{-6}$ & $1.4\times 10^{-3}$ \\
Surface Tension, $\epsilon_{surf}$ [$\rm ergs\,cm^{-2}$] & 200 & \nodata & \nodata & 75	\\
Latent Heat of Vaporization, L [$\rm ergs\,g^{-1}$] & $6.34\times 10^{10}$ &
\nodata & \nodata & $4.87\times 10^{12}$
\enddata
\tablenotetext{a}{We are ignoring here the minor differences between the
densities of the liquid and solid phases of iron and water \citep[CRC Handbook]{CRC:1991}.}
\tablenotetext{b}{The $\rm q_{below}$ parameter is the mole fraction of condensable vapor
just below the cloud base; the calculation of $\rm q_{below}$ assumes 
a solar abundance distribution of the elements \citep{Anders:1989}.}
\tablecomments{Values for the surface tension and latent heat of vaporization are taken from
\citet[CRC Handbook]{CRC:1991}.  They are used to derive $\rm \tau_{nuc}$
using the classical homogeneous nucleation theory and are therefore not applicable to the heterogeneously 
condensing species (see Section \ref{subsection:Microphysical_Timescales}).
Note also that the mole fractions listed apply only to the cloud base (i.e.,
the condensation level, as shown in Figure \ref{figure:Condensation_Level}).  Above
the cloud base, the mole fraction of total condensable 
material in the cloud drops off much more steeply than the gas pressure
(as prescribed in Equations \ref{equation:Rainout_Homogeneous} and \ref{equation:Rainout_Heterogeneous}).  
} 
\end{deluxetable}

In stage (1), two cases must be treated separately, as explained above.  In the
case of homogeneous condensation, we calculate the intersection between the
partial pressure of the condensable vapor and its saturation vapor pressure
curve and identify the condensation level using Equation 
\ref{equation:Condensation_Level}.  In the case of heterogeneous condensation, 
in which the condensing chemical is not present in
vapor form below the cloud base, we determine the cloud base by using the
chemical equilibrium model of \citet{Burrows:1999} (see Figure
\ref{figure:Condensation_Level}).  

In stage (2), we calculate the modal size of particles, which are assumed to
be spherical, at each atmospheric temperature--pressure level.  The model outputs
the modal particle radius.  This radius is determined by growing embryonic
particles of radius $\sim$$10\,\textrm{\AA}$ slowly until the various competing
particle growth timescales are balanced by the timescale of gravitational 
fallout. For the most probable particle radius, the sedimentation timescale
equals the shortest of the growth timescales:
\begin{equation}
\tau_{\rm fall} = \textrm{Min}\,\{\tau_{\rm nuc},\,\tau_{\rm cond},\,
				\tau_{\rm coag},\,\tau_{\rm coal},\,\tau_{\rm conv}\}.
\label{equation:Min_Timescale}
\end{equation}
Note that for homogeneously condensing species in the gas kinetic regime ($\rm Kn >
1$), it may be that radiative cooling, rather than the release of latent heat by
collisions with atmospheric hydrogen, is the dominant process limiting the rate
of particle growth.  Therefore, if the Knudsen number is greater than one
and $\rm \tau^{-1}_{rad} > \tau^{-1}_{nuc}$, we replace $\rm \tau_{\rm nuc}$ in 
Equation \ref{equation:Min_Timescale} with the timescale
for particles to release the latent heat of condensation by radiation, $\rm \tau_{rad}$.  
Additionally, as we have stated in Section \ref{subsection:Microphysical_Timescales},
nucleation does not apply to the case of chemical clouds \citep{Rossow:1978}, and coalescence
is inefficient between solid particles.  The condensed particles can
only accumulate more material by coagulating under the influence of convective updrafts.
Hence, for forsterite and gehlenite, the fall timescale must balance the
minimum of the two timescales relevant for heterogeneously condensing clouds: $\rm
\tau_{coag}$ and $\rm \tau_{conv}$.

The timescales in Equation \ref{equation:Min_Timescale} depend on the Knudsen
and Reynolds numbers (see Section \ref{subsection:Microphysical_Timescales}).
We therefore need a prescription to determine which physical regime applies as the
particles are grown.  Because the Reynolds number depends on velocity, and the
velocity depends on the Reynolds number, calculating the two quantities
independently for a given particle size and atmospheric parameters is a circular problem.
There are no difficulties if the Knudsen number is larger than one, since 
the gas kinetic terminal velocity in that case does not depend on the Reynolds number; the
velocity to use in this case is given by Equation \ref{equation:Gas_Kinetic_Terminal_Velocity}.  
To differentiate between Reynolds number regimes in the small Knudsen number
case (i.e., Equations \ref{equation:Stokes_Terminal_Velocity} and 
\ref{equation:Turbulent_Terminal_Velocity}), we use the following procedure based 
on the fact that each of the two expressions for $\rm v_{fall}$
can be calculated independently of the Reynolds number.  

At each point in the particle growth phase, we use \em both \em 
Equations \ref{equation:Stokes_Terminal_Velocity} and 
\ref{equation:Turbulent_Terminal_Velocity}, along with the known atmospheric
parameters and the current particle size.  In solving for the terminal velocity,
we substitute the SMO surface gravity, $\rm g_{surf}$, in
for g, the local gravitation acceleration.  We then calculate two temporary values
for the Reynolds numbers, one for each of the velocities calculated.  If these
two independently computed Reynolds numbers are both smaller than 
our Reynolds number cutoff point of 70 between laminar and turbulent flow \citep{Rossow:1978}, 
we take the flow around the particles to be 
laminar and adopt the Stokes terminal velocity, Equation \ref{equation:Stokes_Terminal_Velocity}, 
as the actual terminal velocity.  Likewise, 
if they are both larger than the cutoff, we take the flow around the particles to be turbulent and
adopt the turbulent terminal velocity, Equation \ref{equation:Turbulent_Terminal_Velocity}, 
as the actual terminal velocity.  

The intermediate case, in which one temporary Reynolds number is larger than the cutoff but the other 
is smaller than the cutoff, is more difficult to resolve.  We settled on the choice of using 
the lesser of the two terminal velocities as the actual terminal velocity.  
Although this could have been handled in a variety of ways, 
the inherent ambiguities could not have been eliminated without detailed knowledge of the drag coefficient, 
which unfortunately depends in general on the Reynolds number.  Ours was the most conservative choice 
for a first--order estimate of the terminal velocity.  
With this procedure in place to calculate the terminal velocity of particles, 
the effective Reynolds number---which is used to calculate all the other timescales---is 
computed directly from its definition in Equation \ref{equation:Reynolds_Number}, 
with $\rm v_{fall}$ substituted in for the relative velocity, v.

Equation \ref{equation:Min_Timescale} shows the role of convection in this model.  Convection 
directly opposes gravitational sedimentation.  In the presence of
sufficiently vigorous convection, therefore, the particles may continue to accumulate
material until their terminal velocities become equal to the convective updraft velocity.

The simple cloud model delineated above is a first--order estimate
of SMO particle sizes.  The true particle sizes could very well deviate
from the values computed by a factor of order unity, though we believe that 
this factor will be the same for all objects studied because the systematic errors
depend on our assumptions of the unknown parameters of the model, assumptions which we have
applied consistently throughout the computations.  

Our treatment of rainout, in which condensates settle gravitationally and
thus become depleted from the upper atmosphere, differs between homogeneously
and heterogeneously condensing species.  Depletion of the condensate with height must differ
between these two different types of clouds because homogeneous condensation
involves a pressure equilibrium between the vapor and the condensed phase (See
Section \ref{subsection:Microphysical_Timescales}),
whereas heterogeneous condensation produces the condensate directly through
exothermic chemical reactions.  

For the purposes of this discussion, we adopt the notation of \citet{Ackerman:2001} for the relevant
mixing ratios: $\rm q_v =$ moles of vapor per mole of atmosphere, $\rm q_c=$ moles of condensate
per mole of atmosphere.  The total mixing ratio is $\rm q_t=q_v + q_c$.  
We also define $\rm q_{below}$ as the mixing ratio by number (or mole
fraction) of condensable 
vapor just below the cloud base, which equals $\rm q_t$ at that level, since $\rm q_c = 0$ by definition below the cloud 
base (i.e., at levels in the atmosphere where Equation \ref{equation:Condensation_Level} is not satisfied).  
The values of $\rm q_{below}$ for each condensable treated herein are given in Table \ref{table:Condensate_Properties}.

For both homogeneously and heterogeneously condensing species, we assume 
aggressive rainout such that the total material available to condense into particles 
drops off with pressure much more steeply than the gas pressure.  That is, 
at the condensation level, the mixing ratio by number (or equivalently, the mole fraction) 
of the condensing species is taken to be equal to the maximum partial pressure attained by
that species in a gas of solar composition (i.e., parameter $\rm q_{below}$ 
in Table \ref{table:Condensate_Properties}).  For heterogeneously condensing species, 
$\rm q_{below}$ is constrained by the maximum partial pressure of the least
abundant of the constituent molecules combining to form the condensate.  For example, in the case of
forsterite, the magnesium abundance in the gas controls the value of $\rm q_{below}$.  We
have assumed a solar abundance of magnesium in the atmosphere below the
forsterite condensation level.  High above the condensation level, however, the mixing ratio 
of cloud--forming material is significantly reduced from the equilibrium ratio, and the 
cloud becomes more and more tenuous with decreasing pressure.   

For homogeneously condensing species, our rainout prescription is based on the assumption that
all of the supersaturated vapor above the base of the cloud will condense.  As we
have discussed in Section \ref{subsection:Condensation_Level}, the onset of particle
formation requires a supersaturated environment, which thus elevates the cloud base above
the level where precise saturation of condensable vapor is attained.  
Once present, however, particles remain thermodynamically
stable unless the saturation ratio drops to below one, which will not happen
without a temperature inversion in the atmosphere's thermal profile.  For already existent
particles or cloud particle embryos, the vapor in equilibrium with the particles remains exactly at saturation.  
All excess material goes into the condensate.  Hence, throughout the cloud, we take $\rm q_v = P_{sat}/P$, 
where $\rm P_{sat}$ is the saturation vapor pressure and P is the gas pressure.  Notice that this ratio
decreases with altitude because $\rm P_{sat}$ decreases with pressure much
more steeply than the gas pressure, P.  It therefore follows that the partial pressure of the
condensable vapor required to maintain a saturated environment will be
extremely small high above the condensation level.  Though the mole fraction of
condensing vapor, $\rm q_v$, is constrained by the requirement
that the condensed phase remain thermodynamically stable during particle growth, we still need the total
number mixing ratio (i.e., mole fraction) of condensable material, $\rm q_t$, in order to determine $\rm q_c$, the
condensate number mixing ratio.

The character of the rainout itself derives from the assumption 
that the total material available for condensation
above the cloud base follows the saturation vapor pressure and \em not \em the
gas pressure.  We use $\rm q_t = (1 + \mathcal{S}_{max})\,P_{sat}/P$, which implies
that the total condensable material is constrained to drop off with
altitude proportionally to the saturation vapor pressure.  
This is an assumption of the cloud model.  It is not based on a
rigorous derivation of the total mixing ratio with altitude.  The justification for the
rainout scheme is that once condensation begins in the atmosphere, with
subsequent growth and gravitational settling of particles, the condensable material can no longer
remain well--mixed in the atmosphere, as we assume it had been below the cloud base.  
Therefore, the total number of moles of condensable material per mole of gas will
decrease rapidly from its initial value with increasing height above the condensation level.
By contrast with our rainout scheme, a no--rainout model would be one in which the fraction of condensed 
material follows the gas pressure; i.e, $\rm q_t$ equals a constant throughout the cloud.  In this case,
the ratio of condensate to condensable vapor, $\rm q_c/q_v$, will increase
with height above the base of the cloud.

With the above considerations, it is straightforward to calculate $\rm q_c$ as simply the
difference between the total number mixing ratio of condensable material and
the number mixing ratio of condensable vapor: $\rm q_c = q_t - P_{sat}/P$.
We therefore obtain the following equation for $\rm q_c$:
\begin{equation}
\rm q_c = \mathcal{S}_{\rm max}\cdot\frac{P_{sat}}{P} 
= \mathcal{S}_{\rm max}\cdot q_{below}\cdot \frac{P_{c,\,1}}{P}.
\label{equation:Rainout_Homogeneous}
\end{equation}
In Equation \ref{equation:Rainout_Homogeneous}, $\rm P_{c,\,1}$ denotes the saturation vapor pressure
curve---given by $\rm P_{sat}$, as in Equation \ref{equation:Condensation_Level}---adjusted to resemble 
a condensation curve by applying the mixing ratio in a solar composition gas; 
i.e., $\rm P_{sat} = q_{below}\cdot P_{c,\,1}$.  
For example, the dotted curve labeled ``Fe'' in Figure \ref{figure:Condensation_Level} shows 
the $\rm P_{c,\,1}$ for iron, \em not \em the iron saturation vapor pressure curve.  
The subscript notation for $\rm P_{c,\,1}$ simply refers to the ``effective'' 
iron condensation curve as derived from the saturation vapor pressure curve.  
The rainout is expressed in terms of $\rm P_{c,\,1}$ for comparison with the analogous prescription for rainout in a
heterogeneous cloud (Equation \ref{equation:Rainout_Heterogeneous}).  The quantity $\rm P$ in 
Equation \ref{equation:Rainout_Homogeneous} is the total gas pressure in the atmosphere.  As is clear from 
comparing the thick dotted Fe line in Figure \ref{figure:Condensation_Level} 
with one of the gas pressure profiles, $\rm P$ decreases with height above the cloud base much less 
steeply than $\rm P_{c,\,1}$.  The assumption of a constant supersaturation throughout the cloud, 
which we have made in fixing the value of $\rm \mathcal{S}_{max}$, leads to a cloud with a very low ratio of condensed
material to total condensable material unless $\rm \mathcal{S}_{max} \gg 0.01$.

For heterogeneously condensing clouds, we refer to the condensation
curves (denoted $\rm P_{c,\,1}$ in Equations \ref{equation:Rainout_Homogeneous} and \ref{equation:Rainout_Heterogeneous}) 
of forsterite and gehlenite as shown in Figure \ref{figure:Condensation_Level}.  We make the 
assumption for a heterogeneously condensing cloud that the total mixing ratio, $\rm q_{t}$, follows 
the condensation curve above the cloud base.  This assumption is analogous to the assumption used to derive 
Equation \ref{equation:Rainout_Homogeneous} for homogeneously condensing clouds that the total mixing ratio 
follows the saturation vapor pressure curve.  However, in a heterogeneous (or chemical) cloud, 
once the condensate appears in chemical equilibrium, we argue that the solid phase will then be strongly favored, 
and hence the product chemical will be fully condensed.  Thus, for heterogeneous condensates, 
$\rm q_v=0$ and $\rm q_c=q_t$.  The rainout formula for heterogeneous species
is therefore quite similar to Equation \ref{equation:Rainout_Homogeneous},
without the reduction of $\rm q_c$ relative to $\rm q_t$ by the supersaturation factor, which is
not relevant for chemical clouds:
\begin{equation}
\rm q_c = q_{below}\cdot \frac{P_{c,\,1}}{P}.
\label{equation:Rainout_Heterogeneous}
\end{equation}
In Equation \ref{equation:Rainout_Heterogeneous}, the quantity $\rm P_{c,\,1}$ represents the chemical
equilibrium condensation curve of the condensing species, as shown for forsterite and 
gehlenite in Figure \ref{figure:Condensation_Level}.  
It is apparent from Figure \ref{figure:Condensation_Level} that this rainout
prescription sequesters most of the condensate within a scale height of the
cloud base, since the condensation pressure drops off so steeply relative to the gas
pressure.  For example, consider the forsterite cloud shown in Figure
\ref{figure:Condensation_Level} for the brown dwarf model at $\rm T_{eff} = 900\,K$ and 
$\rm g_{surf} = 3\times 10^5\;cm\,s^{-2}$.  At the forsterite cloud base, which 
is at about $\rm T \approx 2000\,K$ and $\rm P \approx 125\,bars$, the mixing
ratio of forsterite is just determined by the abundance of magnesium in the
mixture, or $\rm q_c = q_{below} = 3.2\times 10^{-5}$.  Higher up in the atmosphere (in altitude), at
$\rm T \approx 1700\,K$ and $\rm P \approx 70$ bars, Equation \ref{equation:Rainout_Heterogeneous} 
shows that $\rm q_c$, which follows the condensation curve, is reduced by a factor $\rm \sim$$100$ 
from its value at the cloud base.  However, the gas pressure is down by only about $55\%$, or 
$\sim$$\case{1}{2}$ a scale height.

Equations \ref{equation:Rainout_Homogeneous} and \ref{equation:Rainout_Heterogeneous} show that, 
in the present model, virtually all of the atoms in the solar--composition gas that are initially available for 
condensation become sequestered in the condensate within a scale height of the atmosphere.  For example,
because the limiting atomic species for the formation of forsterite is
magnesium, and forsterite is favored chemically below about 2000 K, magnesium
will have rained out near the forsterite cloud level, and no magnesium will be
available for further condensation into other chemicals high above the
forsterite cloud level.  Rainout thus depletes the atmosphere of the most
refractory elements at progressively lower temperatures and pressures.

Each timescale characterizes the average (or characteristic)
effectiveness of each of the competing microphysical processes in the system.  Our computed particle
sizes thus represent the most probable particle size in the distribution of particle sizes
appearing at each pressure level.  To the order of the accuracy of the particle sizes themselves,
therefore, the resulting particle sizes in each pressure level will equal
the mode of the particle size distribution.  

We make no attempt in the present model to calculate the particle size distribution for each cloud layer.  
Rather, for the purposes of calculating the opacity of clouds and using these
opacities in SMO spectral models, we assume a functional form for the particle size distribution 
that is consistent with measurements of particle size distributions attained in Earth's water clouds 
\citep{Deirmendjian:1964}.  We use the form of the \citet{Deirmendjian:1964} 
exponentially decaying power law employed by \citet{Sudarsky:2000}:
\begin{equation}
\rm n(r) \propto \left(\frac{r}{r_0}\right)^6 exp\left[-6\left(\frac{r}{r_0}\right)\right],
\label{equation:Size_Distribution}
\end{equation}
where n(r)dr is the number of particles per cubic centimeter having radii
$\rm r \to r+dr$.  We employ the cloud model described above to determine the mode
of the distribution, $\rm r_0$.

\subsection{Free Parameters}
\label{subsection:Free_Parameters}

In addition to $\rm \mathcal{S}_{max}$, the assumed maximum supersaturation, the
free parameters of the cloud model include the elemental abundances 
and the sticking coefficients for coagulation and 
coalescence.  The vapor mixing ratios have been calculated assuming solar \citep{Anders:1989} 
abundances of the elements in the atmosphere.

Coagulation and coalescence are not expected to be $100\%$ efficient in all
cases because some molecules stick together when they collide more easily than
others.  Therefore, we have divided the timescales for coagulation and
coalescence, given in \citet{Rossow:1978} as \tcoag and $\tau_{\rm coal}$, by
efficiency parameters $\epsilon_{\rm coag}$ and $\epsilon_{\rm coal}$.

Coalescence between solid particles is extremely inefficient \citep{Rossow:1978}.  
Coalescence will therefore not be important for forsterite and gehlenite because 
the solid phase is strongly favored once the temperature decreases sufficiently for the chemical to appear 
(see Figure \ref{figure:Condensation_Level}).  Therefore, for the heterogeneously condensing materials, 
we have ignored coalescence ($\epsilon_{\rm coal} = 0$).  For the species condensing into 
liquid particles (iron and water), we have taken $\epsilon_{\rm coag}$ = $10^{-1}$ and 
$\epsilon_{\rm coal}$ = $10^{-3}$ \citep{Lunine:1989}.  We have also used 
$\epsilon_{\rm coag}$ = $10^{-1}$ for the coagulation efficiency of heterogeneously condensing species.
Subsequently, we will argue that this choice is relatively unimportant because
coagulation and coalescence generally operate more slowly, for the atmospheres
we are considering here, than heterogeneous nucleation and convection.  We
discuss in Section \ref{subsection:Varying_Free_Parameters} how varying 
these parameters affects the general features of the cloud.

\section{Model Results}
\label{section:Model_Results}

\subsection{Modal Particle Sizes}
\label{subsection:Modal_Particle_Sizes}

We compute particle sizes for each of four species on a set of atmospheric
models spanning the ranges \teff = 600 -- 1600 K and \gsurf
= $1.78\times 10^3$ -- $3\times 10^5\;\textrm{cm s}^{-2}$.  We chose
the most abundant  condensable species in the solar--composition mixture:~iron,
forsterite, and water.  Iron and forsterite are the most
abundant high--temperature condensates \citep{Lunine:1989}.  Although we
have included one of the calcium--aluminum silicates, gehlenite, and these
refractory species do condense into clouds, their abundance in a
solar--composition gas is lower than that of iron or forsterite by a factor of
about ten \citep{Lunine:1989}.  Thus, the contribution of the calcium
and aluminum silicates to the total opacity is expected to be relatively minor.  
Nevertheless, the condensation of the calcium and aluminum silicates is important
for sequestering calcium and aluminum at depth beneath the photospheres of cooler SMOs.  
Moreover, the most refractory calcium and aluminum silicates can potentially serve as
nucleation sites for the condensation of iron vapor via heterogeneous nucleation at low supersaturations.

\subsection{Trends With Brown Dwarf Gravity and Effective Temperature}
\label{subsection:Trends}

\begin{figure}
\includegraphics[height=3.2in, width=2.4in, angle=-90]{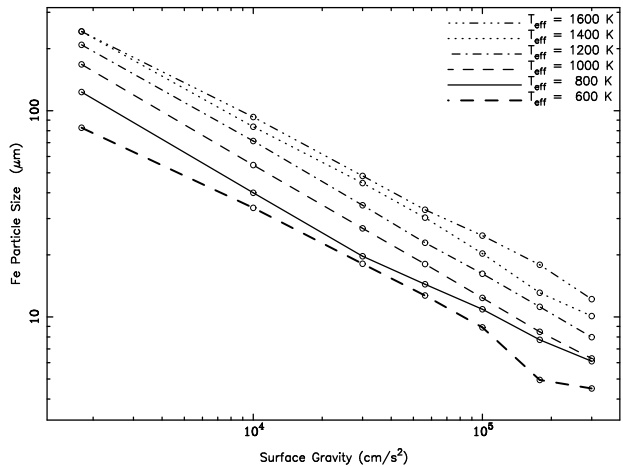}
\caption{
The grain sizes of iron (Fe) at the cloud base for a range of brown dwarf
atmospheric temperature--pressure profiles.  The figure displays the modal
particle sizes obtained by applying the model described in Section \ref{section:Cloud_Model} to each
model brown dwarf atmosphere.  The profiles span a range of effective
temperatures and surface gravities from $\rm {T}_{eff}\,=\,600 - 1600\,K$
and $\rm {g}_{surf}=1.78\times10^3 - 3\times 10^5\;cm\,s^{-2}$.
Each circled point corresponds to a different brown dwarf model atmosphere 
whose effective temperature and gravity depend on the point's location in the
field. The figure shows that high--gravity brown dwarfs will have
characteristically smaller particle sizes than low--gravity brown dwarfs for a
given effective temperature. Similarly, hotter brown dwarfs will exhibit larger
particle sizes, for a given surface gravity, than cooler brown dwarfs.
}
\label{figure:Iron_Sizes}
\end{figure}

\begin{figure}
\includegraphics[height=3.2in, width=2.4in, angle=-90]{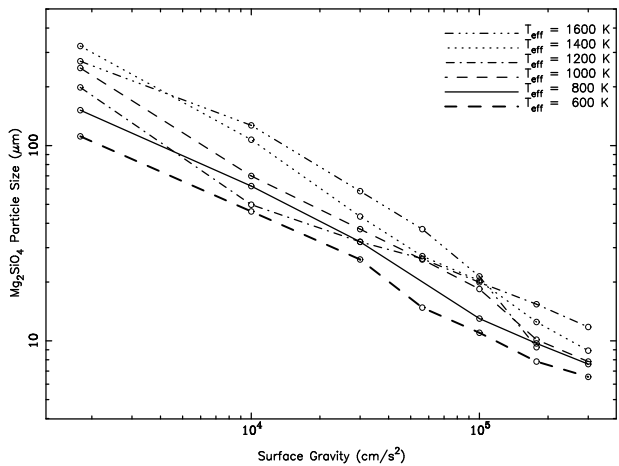}
\caption{
Companion to Figure \ref{figure:Iron_Sizes}:~shows similar trends for forsterite
($\textrm{Mg}_2\textrm{SiO}_4$) particle radii.  The forsterite
particles become extremely large ($\sim$$350\,\micron$) in very low gravity
objects.  Because forsterite forms in abundance in the atmosphere, the optical
effects of forsterite clouds are potentially very important, depending on the
particle sizes.
}
\label{figure:Forsterite_Sizes}
\end{figure}

\begin{figure}
\includegraphics[height=3.2in, width=2.4in, angle=-90]{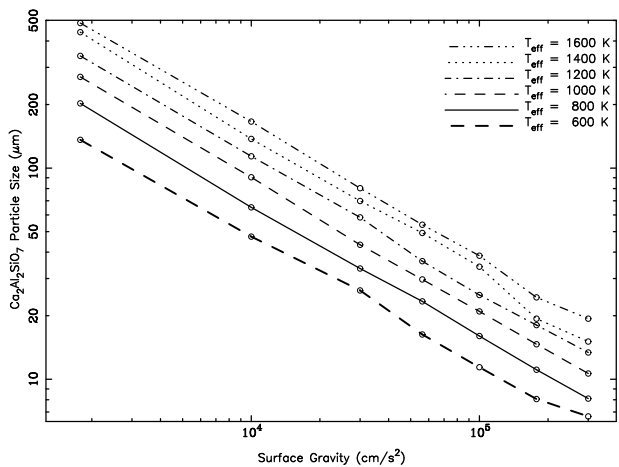}
\caption{
The particle size trends for the calcium--aluminum silicate gehlenite
(Ca$_2$Al$_2$SiO$_7$).  This species is less abundant than forsterite by
about a factor of ten, but it has the important effect of sequestering Ca and Al in
condensed form below the photosphere of the cooler objects.
}
\label{figure:Gehlenite_Sizes}
\end{figure}

Figures \ref{figure:Iron_Sizes}, \ref{figure:Forsterite_Sizes}, and \ref{figure:Gehlenite_Sizes} 
show the particle sizes computed by our model for a range
of brown dwarf atmospheric profiles.  Each circled point represents, for the
atmospheric profile to which it applies, the grain size computed by our model at
the initial cloud level (i.e., the cloud base).  As we discuss further
in Section \ref{subsection:Cloud_Decks_Rad_vs_Conv}, these are in general the
largest modal particle sizes for the cloud, since particle sizes decrease with 
height above the cloud base.  The model atmosphere
corresponding to each circled point is indicated by its position on
the diagram:~gravity increases from left to right, whereas effective
temperature increases from bottom to top in the field.  These graphs say
nothing about the distribution of material above the cloud base.  They
apply only to the initial condensation level where the mass density of material
is greatest.  Figure \ref{figure:Iron_Sizes} shows the results for iron grains, which condense
directly from iron vapor. Figure \ref{figure:Forsterite_Sizes} shows the analogous results for forsterite
grains, which appear as a result of chemistry (i.e., heterogeneous
condensation).  Figure \ref{figure:Gehlenite_Sizes} shows the same for gehlenite, which also condenses
heterogeneously.

Figures \ref{figure:Iron_Sizes}, \ref{figure:Forsterite_Sizes}, and \ref{figure:Gehlenite_Sizes}
show that cloud grains in hot brown dwarfs, for a fixed
surface gravity, are systematically larger than in cold brown dwarfs.
Similarly, cloud particles in objects with high surface gravities, for a fixed
effective temperature (e.g., older brown dwarfs), are systematically smaller
than in objects having low surface gravities.

The trend with surface gravity is simple to explain:~particles settle more
quickly in high--gravity environments.  This appears to be the overriding
effect governing particle size.  Although both the atmospheric pressure and
temperature at the cloud base are higher in high--gravity objects, and the
convection therefore more rapid, gravity dominates and the resulting
particles are smaller.  For the lowest gravity objects, the forsterite particles
grow quite large.  They therefore contribute only negligibly to the total optical
depth \citep{Sudarsky:2000}.

The trend with effective temperature results from the difference in convective
updraft velocity between colder and hotter objects.  Objects with higher
effective temperatures must transport a higher flux of thermal energy to the
surface via interior turbulent convection.  Therefore, the updrafts holding
cloud particles aloft will be more rapid in the higher $\textrm {T}_{\rm eff}$
objects, resulting in larger particle sizes.

\subsection{Cloud Decks in Convective vs. Radiative Regions}
\label{subsection:Cloud_Decks_Rad_vs_Conv}

Brown dwarfs are almost fully convective, with only a thin radiative atmosphere
\citep{Basri:2000,Burrows:2001}.  Convective uplifting sustains the particles 
against gravitational fallout.  The effectiveness of
this process depends on the velocity scale of the convection.  Hence, the
gradient in the convective velocity plays a role in determining particle size.
Deep in the atmosphere, the brown dwarf is fully convective, with a relatively
high updraft velocity (approaching $\rm 5\times 10^3\;cm/s$).  At higher altitudes, the updraft
speed is small, but \em not \em zero.  Therefore, clouds at depth are convective; clouds at
altitude are quiescent.  By quiescent, we mean clouds formed in convectively 
stable layers where the true lapse rate is shallower than the adiabatic lapse
rate.

The updraft speeds observed in stably stratified layers in Earth's atmosphere 
typically range from $\rm 1-10\;cm/s$.  Clouds analogous to Earth's stratus clouds are 
possible in quiescent layers of planetary atmospheres, even though the updraft 
speeds are smaller by a factor of a hundred or more than the typical speeds 
of convective updrafts \citep{Rogers:1989}.  In the case of these stratiform 
clouds, small particle sizes (generally less than $\rm 10\,\mu m$) lead to
relatively slow sedimentation rates.  Large--scale uplifting of a
stable layer can thus replenish the cloud material lost by evaporating particles on a 
timescale of hours or days, which is rapid enough to sustain the cloud.  
Slow, large--scale updrafts, which we assume are occurring in these atmospheres,
are too slow to affect the particles directly.  Therefore,
we do not include them as a growth process in the cloud model, as we have done
for convective updrafts, but their existence is an important feature 
of the formation of quiescent clouds by maintaining the supersaturated 
environment necessary for particle growth.

\begin{figure}
\includegraphics[height=3.7in, width=2.8in, angle=-90]{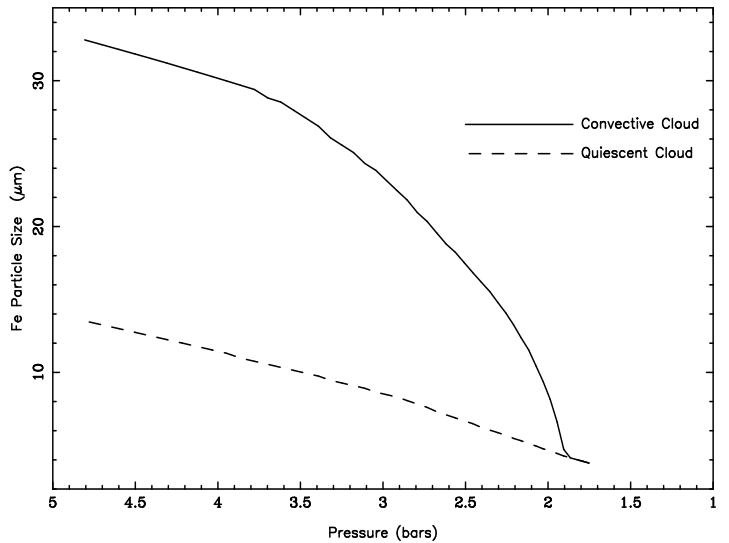}
\caption{
The structure of an iron cloud deck for a particular brown dwarf at
$\textrm {T}_{\rm eff}=1500$ K and $\textrm {g}_{\rm surf} = 5.62\times 10^4\;
\textrm{cm s}^{-2}$, assuming a constant $1\%$ maximum supersaturation during the
entire particle growth phase.  The dashed curve assumes that convection is not
effective at sustaining particle growth.  Therefore, the particles do not grow
as large.  For this radiative (or quiescent) cloud, the particle sizes vary by
about a factor of two within an atmospheric scale height.  On the other hand, for the convective
cloud, in which we allow updrafts to sustain particle growth against
gravitational sedimentation, the particles grow relatively large.  Because the
cloud base is slightly above the radiative--convective boundary of the profile,
the updraft velocity decreases rapidly with height. Thus, for clouds forming
near unity optical depth, we expect variable particle sizes from the cloud base to
the cloud top.  The cloud particle sizes for the convective cloud merges
with the sizes obtained for the quiescent cloud once the updraft velocity becomes
negligible.  The exact cutoff for the cloud is somewhat arbitrary; 
we present a full scale height of both cloud decks.  As we explain in our discussion of rainout in
Section \ref{subsection:Cloud_Code}, the cloud is extremely thin this high above the cloud base,
and the mole fraction of total condensing material, given by $\rm q_t$, is depleted from its value at 
the cloud base, where $\rm q_t = q_{below}$, by many orders of magnitude 
(see Equation \ref{equation:Rainout_Homogeneous}).
}
\label{figure:Cloud_Decks_Rad_vs_Conv}
\end{figure}

Particles in quiescent clouds are generally smaller than particles in convective
clouds.  Figure \ref{figure:Cloud_Decks_Rad_vs_Conv} compares the particle sizes of condensed iron for a
brown dwarf model having \teff = 1500 K and \gsurf =
$5.62\times10^4\;\textrm{cm s}^{-2}$.  The modal particle sizes
in the cloud are shown for iron under two contrasting assumptions: (1) the cloud
is convective (i.e., $\rm \tau_{conv} = H/v_{conv}$), and (2) the cloud is
radiative (i.e., $\tau_{\rm conv} = \infty$).  Our calculations suggest that for
this atmosphere, a convective iron cloud is more realistic.  We thus
predict the solid curve for the actual particle size distribution in the cloud
deck.  Figure \ref{figure:Cloud_Decks_Rad_vs_Conv} compares the particle sizes obtained by activating and
deactivating the convective uplifting mechanism.

Figure \ref{figure:Cloud_Decks_Rad_vs_Conv} shows, for this particular brown dwarf 
atmosphere, that cloud particles growing in a region of strong convective upwelling 
are largest near the cloud base but then diminish rapidly with height. 
This decrease in size is due to the proximity of the iron cloud base, for this particular model, to the
radiative--convective boundary of the atmosphere, where the gradient in the
upwelling velocity is steep.  For convective clouds deep within the convective
zone, where the upwelling velocity is nearly constant, the particle sizes will be
more constant throughout the cloud deck.  However, deep clouds will not strongly
influence the object's emergent spectrum.

If the cloud is radiative, the particles will not grow as large as they would
under the influence of convection.  Furthermore, their sizes will not
decrease by more than a factor of two from base to cloud top.  This result
appears quite general for the radiative clouds we have investigated.

The structure of the cloud decks (e.g., as shown in Figure \ref{figure:Cloud_Decks_Rad_vs_Conv}) 
of brown dwarfs is crucial for computing the optical depth of the cloud.  For clouds in which
the overall particle sizes decrease significantly with altitude above the cloud
base, it may be that the bulk of the optical depth in the near--infrared will be
contributed not by the lowest cloud layers but by intermediate layers having
significant abundances of smaller particles.

\subsection{Varying the Free Parameters}
\label{subsection:Varying_Free_Parameters}

Among the model free parameters discussed above, the level of maximum
supersaturation is the most problematic.  The value of \smax is particularly
difficult to calculate independently, and because the condensation timescale 
depends on it, \smax has a significant effect on the resulting
particle size.  This effect will be manifest in quiescent clouds but will be
less important in convective clouds.  This is because in rapidly convecting regions,
convective upwelling can directly sustain particle growth.

\begin{figure}
\includegraphics[height=3.7in, width=2.8in, angle=-90]{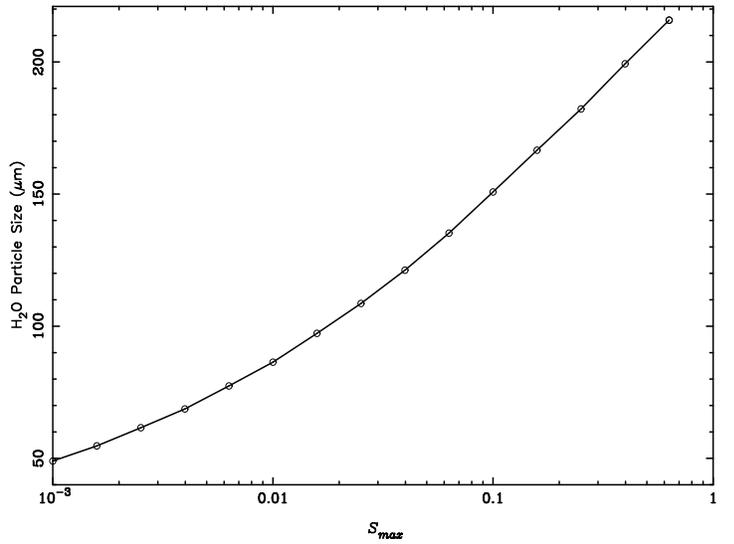}
\caption{
The modal particle sizes, as a function of $\mathcal {S}_{\rm max}$, for a water
cloud in the upper atmosphere of a cool brown dwarf: $\textrm {T}_{\rm eff}=350$ K
and $\textrm {g}_{\rm surf} = 10^4\;\textrm{cm s}^{-2}$.  Convection is not
occurring in these outer layers.  The figure demonstrates the difficulty in estimating
particle sizes for radiative (i.e., quiescent) clouds because of the strong dependence of particle size
on the assumed maximum supersaturation.  Water cloud particles will grow extremely large in a
clean atmosphere; i.e., one in which the paucity of condensation nuclei requires
high levels of supersaturation of the vapor in order for nucleation to begin.
In this case, nucleation proceeds rapidly and the particles grow very large.
For convective clouds, on the other hand, the value assumed for the maximum
supersaturation is not so important, as the cloud particles will grow to nearly the
same size for any value of \smax $< 100\%$.
}
\label{figure:Particle_Sizes_vs_Smax}
\end{figure}

Figure \ref{figure:Particle_Sizes_vs_Smax} shows, for one brown dwarf model, the effect of varying
\smax on the modal particle size of a water cloud.  For low levels
of supersaturation, the particles grow to a radius of $\sim$50 $\mu
\textrm {m}$.  In this case, abundant condensation nuclei raise the rate of
heterogeneous nucleation high above the rate of homogeneous nucleation, and
condensation can begin as soon as the atmosphere becomes slightly supersaturated.  For high
levels of supersaturation ($>$$10\%$), they grow much larger, to $\sim$$150\,
\textrm {\micron}$. In this clean atmosphere case, heterogeneous nucleation does
not occur, and the partial pressure of water in the atmosphere must greatly
exceed the saturation vapor pressure in order for condensation to begin.  We
have assumed an intermediate value of \smax$=10^{-2}$ for iron.
This value is our best estimate for the \smax of vaporous iron based on
a knowledge of the maximum supersaturations attained in terrestrial water
clouds \citep{Rossow:1978}.

The other free parameters of the cloud model are less difficult.  The mixing
ratios depend on the metallicity, which we have taken to be solar.  The
chemical equilibrium code requires the metallicity to be specified.  Once the
assumption of elemental composition is made, however, we can calculate the
partial pressures of all the gases in the mixture.  We defer a discussion of the
effect of brown dwarf metallicity on cloud structure to future work.

The efficiency parameters for coagulation and coalescence, $\epsilon_{\rm coag}$ and
$\epsilon_{\rm coal}$, are potentially important.  However, our model results
show that the timescales for these processes, whatever their efficiency, are
consistently longer than the timescale of growth by heterogeneous nucleation at 
1$\%$ supersaturation.  The model suggests that coalescence is much faster
than coagulation for the two species---iron and water---in which coalescence
operates efficiently.  The coalescence timescale, for the largest particles we have grown
($\sim$500 $\mu \textrm {m}$), can approach the timescale of heterogeneous 
nucleation. However, for the clouds in most of the atmospheres we have treated,
coalescence is still slower than heterogeneous nucleation by a factor of ten or
more. Indeed, unless the efficiency of coalescence is greater than one,
requiring charged aerosols \citep{Rossow:1978}, coalescence will never
dominate particle growth.

\section{Comparison With Other Cloud Models}
\label{section:Cloud_Model_Comparisons}

\subsection{Ackerman \& Marley (2001)}
\label{subsection:Ackerman_Marley_model}

\citet{Ackerman:2001} also employ a one--dimensional cloud model in which cloud layers
are treated as horizontally uniform; i.e., in a globally averaged sense.  
The \citet{Ackerman:2001} cloud model arrives at particle sizes by balancing
the upward transport of condensable material due to convective updrafts with the
downward sedimentation of condensate.  In the case in which the convective
timescale, $\rm \tau_{conv}$, dominates over the other growth timescales in the present
cloud model, which will occur in vigorously convecting regions of the atmosphere, the
results of the present cloud model are expected to produce very similar
results to the \citet{Ackerman:2001} treatment.  This is because our formulation 
then reduces to a similar comparison between the characteristic velocities of 
turbulent convective updrafts and the terminal velocity of particles.  

In their treatment of rainout, \citet{Ackerman:2001} 
introduce an additional parameter, $f_{\rm rain}$, which characterizes the vertical distribution of
condensate in the cloud.  The $f_{\rm rain}$ parameter, as 
\citet{Ackerman:2001} state, is difficult to calculate from basic principles.  They leave it as
an adjustable parameter in their model.  It depends on the mass--weighted
sedimentation velocity of the cloud droplets and the convective velocity scale.
The difficulty in calculating $f_{\rm rain}$ lies in the complexity of modeling
fully the nature of the convection within the cloud.  This problem has yet to be
successfully attacked for brown dwarf atmospheres.

If it can be computed for these convective clouds, however, knowledge of
$f_{\rm rain}$ can potentially provide a more realistic measure of the height and
distribution of material in the cloud.  For quiescent clouds, $f_{\rm rain}$ does
not play a role.  \citet{Ackerman:2001} have assumed $f_{\rm rain}$ = 2--3 in
their calculations.  The cloud decks they derive are somewhat more compact
than those produced in the present cloud model, although both prescriptions
lead to rapid depletion of cloud material within a fraction of an atmospheric
scale height above the condensation level.

The particle sizes we obtain compare favorably with the sizes predicted by the
\citet{Ackerman:2001} model.  They predict modal particle radii in the
intermediate size range of 40-80 $\mu \textrm {m}$ for both iron and silicate
grains, in good agreement with the particle size ranges shown in 
Figures \ref{figure:Iron_Sizes}, \ref{figure:Forsterite_Sizes}, and \ref{figure:Gehlenite_Sizes}.

\subsection{Helling \em et al. \em (2001)}
\label{subsection:Helling_model}

\citet{Helling:2001} study the onset of cloud particle growth via
acoustic waves.  They show that small ($1-10\,\micron$ in radius) sized particles can
nucleate rapidly, normally within a few seconds.  These values are consistent
with our particle sizes in radiative regions under the assumption of very low
supersaturations ($\mathcal{S}_{\rm max} \ll 10^{-2}$).  Our larger particles form from
sustained particle growth by convective uplifting. The grains generated in 
\citep{Helling:2001} are not grown to the maximum size allowed gravitationally, 
which we assume can occur.  

The predictions of the two models in terms of particle radii will be in qualitative 
agreement in the absence of the $\rm \tau_{conv}$ timescale employed in our model. 
Nevertheless, we emphasize caution in comparing directly the results of these two
models.  They employ vastly different physical approaches and are therefore
not expected to agree in many cases, even qualitatively.  The model of
\citet{Helling:2001} attempts a detailed simulation of particle growth in brown
dwarf atmospheres, including the complicated effects of hydrodynamics.  
The goal of the present cloud model, rather, is not to directly simulate
the dynamics of particle growth but to develop a computationally economical cloud model 
that can be incorporated easily into spectral synthesis models.

\section{Discussion}
\label{section:Discussion}

\subsection{Cloud Opacity}
\label{subsection:Cloud_Opacity}

The effects of clouds on the emergent spectra of substellar objects
depend strongly on grain sizes.  We derive the wavelength--dependent
absorption and scattering opacities of grains with a full Mie theory
approach, utilizing the formalism of \citet{vandeHulst:1957} and
\citet{Deirmendjian:1969}.

\begin{figure}
\includegraphics[height=3.7in, width=2.8in, angle=-90]{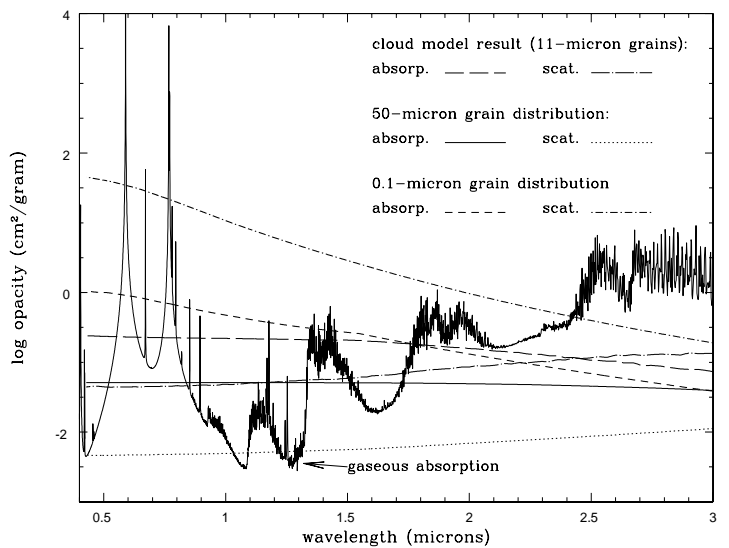}
\caption{
Scattering and absorption opacities of forsterite are compared with
gaseous atomic and molecular absorption at (1600 K, 2.5 bar).  Our modeled modal particle
size, $11\,\micron$, in a brown dwarf (T$_{\textrm{eff}}$ = 1500 K, $\rm g_{surf}$ =
$10^5$ cm s$^{-2}$) is contrasted with both a larger size distribution peaked at
$50\,\micron$ and a smaller size distribution peaked at $0.1\,\micron$
(representative of an interstellar grain size assumed by some researchers).  
In all cases, a functional form of the distribution about the modal size is 
used \citep{Deirmendjian:1964}.
}
\label{figure:Cloud_Opacity}
\end{figure}

Figure \ref{figure:Cloud_Opacity} shows the results of such a calculation for forsterite grains in a
brown dwarf atmosphere (T$_{\rm eff}$ = 1500 K, g$_{\rm surf}$ = $10^5$ cm
s$^{-2}$) based on our cloud model results and optical constants from \citet{Scott:1996}.  
Also shown are the results for a larger grain size distribution
peaked at 50 $\rm \mu m$, as well as for a size distribution peaked at 0.1
$\rm \mu m$, which is representative of an interstellar particle size \citep{Mathis:1977} 
often assumed to be appropriate for substellar objects \citep{Allard:2001,Barman:2001}. 
For comparison, we have plotted the atomic
and molecular gaseous opacities \citep[and references therein]{Burrows:2001} within the cloud region.  
The substantial absorption and scattering differences between the size
distributions, and relative to the gaseous absorption, convey the importance of
proper cloud modeling.

\subsection{Effect on Spectra}
\label{subsection:Spectral_Effects}

\begin{figure}
\includegraphics[height=3.2in, width=2.4in, angle=-90]{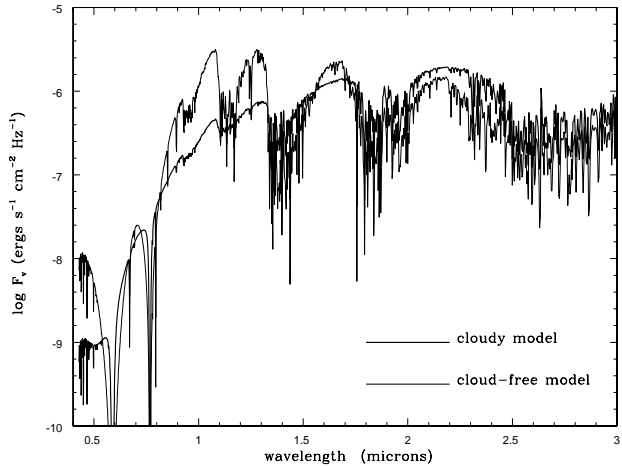}
\caption{
The emergent spectrum of a cloudy brown dwarf (T$_{\textrm{eff}}$
= 1500 K, $\rm g_{surf}$ = $10^5$ cm s$^{-2}$) is compared with a cloud--free model
of the same effective temperature and surface gravity.  In the cloudy model, the
base of the forsterite cloud resides at approximately 4 bars (1800 K).
For the synthetic cloudy spectrum shown, we took the 
cloud deck to be a scale height thick.  This choice will not affect the shape
of the model spectrum, however, so long as the cloud deck is more than about a third of a scale height thick, 
since the cloud is very tenuous and thus optically thin high above the cloud base.  The particle
size distribution was chosen to be the same throughout the cloud deck for
simplicity.  We used a \citet{Deirmendjian:1964} distribution centered around $\rm 11\,\mu m$, which was the typical
particle size in the forsterite cloud deck as computed by the cloud model.
Within the cloud region, the full solar abundance of magnesium has been condensed into forsterite grains.
}
\label{figure:Model_Spectra}
\end{figure}

Figure \ref{figure:Model_Spectra} depicts the effects of our modeled forsterite cloud on
the emergent spectrum of a brown dwarf (\teff = 1500 K,
\gsurf = $10^5$ cm s$^{-2}$).  This model atmosphere was obtained using the
self--consistent stellar atmosphere code, TLUSTY \citep{Hubeny:1988,Hubeny:1995}.  The base of the forsterite
cloud resides at approximately 4 bars and 1800 K, which is the highest temperature in this atmosphere at
which forsterite grains can form.  

We made two simplifying approximations in producing the cloudy model spectrum shown 
in Figure \ref{figure:Model_Spectra}.  First, a cutoff for the 
cloud deck was desirable to facilitate the calulation of optical depth.  
We chose this cutoff to be one scale height above the cloud base, 
which is the atmospheric pressure e-folding distance.  Since the remaining cloud--forming material 
is highly depleted of condensate a scale height above the condensation level 
(see Section \ref{subsection:Cloud_Code}, Equation \ref{equation:Rainout_Heterogeneous}), 
the cloud in this region is optically thin.  This simplification is therefore not a
concern, and the spectral model shown in Figure \ref{figure:Model_Spectra} 
incorporates the full opacity of the cloud.

Second, owing to difficulties in computing the Mie theory scattering and 
absorption opacities iteratively, we chose to simplify the radiative transfer problem by
using a uniform modal particle radius of $\rm 11\,\mu m$ throughout the cloudy region.  
That is, at every atmospheric pressure level for which we computed the opacity
of particles, we employed the \citet{Deirmendjian:1964} particle size distribution, 
given by Equation \ref{equation:Size_Distribution}, with value of $\rm r_0$ equal to $\rm 11\,\mu m$.  
The scattering and absorption opacities of a cloud of $\rm 11\,\mu m$ forsterite particles are shown in 
Figure \ref{figure:Cloud_Opacity}.  This value represents the typical particle size in the 
cloud deck, accounting for the decrease in density of cloud 
material with height above the cloud base (i.e., a number density weighted average of the particle sizes 
predicted by the cloud model).

A cloud--free model of the same effective temperature and gravity is plotted for comparison.  
Within the $B$ ($\sim$ 0.45 $\mu$m) and $Z$ ($\sim$ 1 $\mu$m) bands, the emergent flux is lowered by as much
as one dex. However, the strong absorption by the wings of the sodium and
potassium resonance lines is mitigated somewhat due to the clouds.  Also of
interest are the differences in the $J$ ($\sim 1.25 \, \mu$m), $H$ ($\rm \sim 1.6 \, \mu
m$), and $K$ ($\rm \sim 2.2 \, \mu m$) infrared bands.  The presence of clouds reduces
the emergent flux in the otherwise relatively clear $J$ and $H$ bands, allowing
more flux to escape between these bands and at longer wavelengths.  In fact, the
$J-K$ colors differ by over 1.5 magnitudes between the cloudy and cloud--free
models.

The cloudy spectrum presented in Figure \ref{figure:Model_Spectra} 
is intended to demonstrate the potential importance of clouds as an opacity source in
SMO atmospheres.  We defer to future work the problem of more realistically 
incorporating the opacity of clouds into spectral synthesis models.

\subsection{Coupling Clouds With SMO Atmosphere Models}
\label{subsection:Coupling_Clouds_With_SMO_Atmosphere_Models}

Developing fully self--consistent atmosphere models with clouds is potentially 
problematic because clouds perturb the radiative balance of
the upper atmosphere.  Scattering and absorption of radiation by the cloud causes 
the temperature--pressure structure of a cloudy atmosphere to deviate 
from the structure of a cloud--free atmosphere having the same effective temperature and surface gravity.  
The extent of the effect is strongly dependent on the location of the cloud in the 
atmosphere and the vertical variation of cloud particle sizes and number densities.  
Calculating an atmospheric profile that is in radiative equilibrium including both the gaseous
opacity and the opacity of the cloud is straightforward.  The potential inconsistency arises 
from the fact that the nature of the cloud itself, including the cloud base location and the particle sizes,  
depends in general on the temperature--pressure profile of the atmosphere, 
as we show in Section \ref{section:Cloud_Model}.  This is a problem faced by all 
research groups attempting to include clouds completely self--consistently 
into SMO spectral synthesis models.


We explored this problem quantitatively by comparing the temperature--pressure
profile of two atmosphere models, a ``cloudy'' model atmosphere and a
``cloud--free'' model atmosphere, at the same effective temperature and surface
gravity.  The cloud--free model used for this test was obtained using 
TLUSTY, our self--consistent model atmosphere code \citep{Hubeny:1988,Hubeny:1995},
by including only the opacity from 
gaseous atomic and molecular absorption.  We then applied the cloud model 
described in Section \ref{section:Cloud_Model} to this cloud--free
atmosphere to obtain the distribution of particle sizes and densities of
the forsterite cloud predicted to form near the object's visible surface.  
For simplicity, we took a sensible average of the particle sizes output by the 
cloud model and calculated, for a cloud of uniform modal particle size, 
the absorption and scattering of radiation by the particles 
using the Mie theory approach outlined in Section \ref{subsection:Cloud_Opacity}.  We 
then incorporated this opacity back into the atmosphere code to recompute the
temperature--pressure structure, thus obtaining the cloudy model atmosphere.  
We found the cloudy atmosphere to be hotter by several hundred degrees 
than the cloud--free atmosphere at the same pressure, a significant change.
This is the back-warming effect of the cloud alluded to previously.

We then applied the cloud model to the new cloudy atmosphere to see
whether the forsterite grain sizes varied significantly as a result of the
change in the temperature--pressure profile.  We found that the particles
did change in size:~their radius increased by about a factor of three.  That is, the
forsterite cloud looks quite different for the perturbed atmosphere model
than it does for the original cloud--free model.  This change is a result
of the fact that the forsterite cloud straddles the convective--radiative
boundary of the atmosphere.  Thus, in the original model, the forsterite cloud
appears in a convectively stable region of the brown dwarf, but in the perturbed model, 
the forsterite cloud forms in the convective zone of the atmosphere.  Thus,
the particles in the perturbed model came out larger than the particles
in the original model (Figure \ref{figure:Cloud_Decks_Rad_vs_Conv} shows
how particles in radiative regions are systematically smaller than particles
forming in convective regions).

We performed a similar test for iron clouds.  Unlike for the forsterite cloud, 
we found only a small change in the particle sizes---a decrease of about $15\%$ in 
radius---resulting from the perturbation in the atmospheric temperature--pressure structure.  
The iron cloud forms deeply enough in the convective region of the brown dwarf 
studied to not be strongly affected by the heating due to the cloud.
The degree of inconsistency between cloud--free and cloudy profiles therefore depends on the details 
of the calculation itself.  In many cases, the discrepancies will not be a major concern, but
they can be particularly large if the condensation curve happens to intersect
the atmospheric profile near the convective--radiative boundary. 

We emphasize that both the ``cloud--free'' and ``cloudy'' model atmospheres described
above satisfy the radiative equilibrium boundary condition, and in that sense, the
profiles are self--consistent.  The remaining inconsistency arises from not accounting
properly for possible adjustments to the opacity of the cloud when the particle sizes change
after the profile is perturbed.  A future challenge will be developing the machinery 
to generate model atmospheres that incorporate clouds fully self--consistently.

\section{Conclusions}
\label{section:Conclusions}

\subsection{General Results}
\label{subsection:General_Results}

We have addressed the condensation and subsequent growth of
cloud particles in the atmospheres of brown dwarfs.  We present optimal
particle sizes for three abundant species---iron, forsterite, and
gehlenite---for a broad range of brown dwarf effective temperatures and surface
gravities.

High-gravity brown dwarfs exhibit clouds with typical particle sizes in
the $5-20\,\micron$ range.  The particles grow much larger, however, in
low-gravity objects, often greater than $100\,\micron$ in radius.  We discovered
a similar trend with effective temperature:~hot brown dwarfs have
characteristically larger particle sizes than cool brown dwarfs because of the
increased energy flux that must be transported by convection.  The distribution
of cloud particle sizes depends strongly on the atmospheric parameters, and it
is therefore unrealistic in spectral models to assume a single particle size
distribution for the entire class of SMOs.

We demonstrate that particle size crucially affects the optical depth of the
cloud.  Unlike clouds having a particle size distribution centered at
$0.1\,\micron$, these cloud decks do not dominate the opacity.  Rather, they
smooth the emergent spectrum and partially redistribute the radiative energy
(see Figure \ref{figure:Model_Spectra}).

\subsection{Improved Cloud Models: Morphology and Patchiness}
\label{subsection:Improving_Cloud_Models}

A complete theory of brown dwarf and giant planet atmospheres will
require detailed modeling of clouds.  Although obtaining plausible particle
sizes is a good start, we have been unable to say anything yet about the
morphology or patchiness of clouds.  For example, on Jupiter, clouds appear
in banded structures that vary latitudinally, while Neptune and Uranus show less
latitudinal banding in their surface cloud patterns.  The structure and
patchiness of clouds on brown dwarfs is not known, though the putative
variability reported by \citet{Bailer-Jones:2001} suggests that cloud
patches are not necessarily static.

A simple approach to understanding surface cloud distributions is the moist
entraining plume model.  Such a model has been put forth by \citet{Lunine:1989} 
based on the work of \citet{Stoker:1986} for Jupiter's equatorial
plumes.  The model generates plumes from atmospheric material whose buoyancy is
increased by the release of latent heat from condensation.

More elaborate calculations would involve three--dimensional modeling
of the general atmospheric circulation.  Such an approach would be quite difficult to implement, especially without
accurate meteorological data for SMO atmospheres.  General circulation models would also be too 
intensive computationally to directly incorporate into SMO spectral models.

\subsection{Future Work in Spectral Synthesis}
\label{subsection:Future_Work}

The results of the current work will be useful in developing more elaborate
spectral models of substellar atmospheres.  We plan to follow up the present
work with an exploration of the spectral effects of a variety of cloud
compositions and distributions.  Although Figure \ref{figure:Model_Spectra} shows the general effects of
introducing a forsterite cloud on the spectrum of a brown dwarf, we have not yet
varied the particle size distribution from the base to the top of the cloud.  A
large drop in the particle sizes with height above the cloud base will produce
greater optical depth than a cloud deck having uniformly large particles.  It is
also likely that clouds of more than one species will form near the photosphere
simultaneously.  A major future challenge of this field will be to model the time dependent
dust formation processes of all the major condensable species together, 
including the possible formation of \em core--mantle \em grains, which are composed 
of multiple chemical phases, and then to incorporate them fully self--consistently 
into substellar model atmospheres.

\acknowledgements

We thank Ivan Hubeny, Chris Sharp, Jonathan Fortney, Jason Barnes, Adam Showman, Bill Hubbard, 
Jim Liebert, and the referee for insightful discussions and advice. This work
was supported in part by NASA grants NAG5-10450, NAG5-10760 and NAG5-10629.


\end{document}